\documentclass{jpp}
\usepackage{bm}
\usepackage{epsfig}
\usepackage{graphicx}
\usepackage{xcolor}
\usepackage{braket}

\usepackage[utf8]{inputenc}
\usepackage[T1]{fontenc}
\usepackage{amsmath}
\usepackage{subfig}
\usepackage{flexisym}
\usepackage{wasysym}
\usepackage{natbib}
\usepackage{enumerate}   
\usepackage{amssymb}

\usepackage[final]{hyperref} 
\hypersetup{
	linkcolor=blue,        
	citecolor=blue,        
	filecolor=magenta,     
	urlcolor=blue         
}
\defcitealias{zhou2019magnetic}{Z19}

\newcommand{\be}{\begin{equation}}
\newcommand{\ee}{\end{equation}}
\newcommand{\beq}{\begin{equation}}
\newcommand{\eeq}{\end{equation}}

\newcommand\bea{\begin{eqnarray}}
\newcommand\eea{\end{eqnarray}}

\shorttitle{Multi-scale dynamics of magnetic flux tubes and inverse magnetic energy transfer}
\shortauthor{Muni Zhou, Nuno F. Loureiro and Dmitri A. Uzdensky}

\title{Multi-scale dynamics of magnetic flux tubes and inverse magnetic energy transfer}
\author{Muni Zhou}
\affiliation{Plasma Science and Fusion Center$,$ Massachusetts Institute of Technology$,$ Cambridge$,$ MA 02139$,$ USA}
\author{Nuno F. Loureiro}
\affiliation{Plasma Science and Fusion Center$,$ Massachusetts Institute of Technology$,$ Cambridge$,$ MA 02139$,$ USA}
\author{Dmitri A. Uzdensky}
\affiliation{Center for Integrated Plasma Studies$,$ Physics Department$,$ UCB-390$,$ University of Colorado$,$ Boulder$,$ CO 80309$,$ USA}

\author{Muni Zhou\aff{1},
  Nuno F. Loureiro\aff{1},
 Dmitri A. Uzdensky\aff{2}
}

\affiliation{\aff{1}Plasma Science and Fusion Center$,$ Massachusetts Institute of Technology$,$ Cambridge$,$ MA 02139$,$ USA
\aff{2}Center for Integrated Plasma Studies$,$ Physics Department$,$ UCB-390$,$ University of Colorado$,$ Boulder$,$ CO 80309$,$ USA}

\begin{document}

\maketitle

\begin{abstract}
We report on an analytical and numerical study of the dynamics of a three-dimensional array of identical magnetic flux tubes in the reduced-magnetohydrodynamic description of the plasma. 
We propose that the long-time evolution of this system is dictated by flux-tube mergers and that such mergers are dynamically constrained by the conservation of the pertinent (ideal) invariants, {\it viz.} the magnetic potential and axial fluxes of each tube. We also propose that in the direction perpendicular to the merging plane, flux tubes evolve in critically-balanced fashion. These notions allow us to construct an analytical model for how quantities such as the magnetic energy and the energy-containing scale evolve as functions of time. Of particular importance is the conclusion that, like its two-dimensional counterpart, this system exhibits an inverse transfer of magnetic energy that terminates only at the system scale.
We perform direct numerical simulations that confirm these predictions and reveal other interesting aspects of the evolution of the system. We find, for example, that the early time evolution is characterized by a sharp decay of the initial magnetic energy, which we attribute to the ubiquitous formation of current sheets.
We also show that a quantitatively similar inverse transfer of magnetic energy is observed when the initial condition is a random, small-scale magnetic seed field.
\end{abstract}

\section{Introduction}
Inverse transfer of magnetic energy from small to large scales in highly conducting plasmas is an important topic in modern plasma physics and astrophysics. 
It is particularly important for understanding magnetogenesis~\citep{kulsrud2008origin}, i.e., the origin of large-scale magnetic fields that are ubiquitous in space and astrophysical systems. 
Examples of such systems include planets, such as Earth and Jupiter, main-sequence stars including the Sun, neutron stars and white dwarfs, accretion disks around black holes, the interstellar medium (ISM) in galaxies including the Milky Way, and, on even larger scales, the hot gas in the intracluster medium (ICM).
While the leading paradigm for the origin of large-scale fields invokes turbulent magnetohydrodynamic (MHD) $\alpha\Omega$ dynamo~\citep{Parker1955dynamo},
driven by turbulence coupled with large-scale differential rotation~\citep{Goldreich1965}, 
successful dynamo action needs a seed magnetic field of sufficient strength and sufficiently large coherence length to overcome resistive (or other nonideal) losses. 
In some natural systems, however, the seed field may originate at very small, kinetic scales and then has to traverse  many orders of magnitude in coherence length to reach the desired, extremely large, astronomical scales.

An important example of this scenario arises when the initial field is generated via the 
Weibel instability~\citep{weibel1959,burton1959} --- for example, in a collisionless shock~\citep{medvedev1999generation}. 
This is relevant to systems as diverse as gamma-ray-burst (GRB) jets, supernova explosions, and large-scale accretion shocks at the outskirts of galaxies during early stages of galaxy formation~\citep{gruzinov2001gamma}. 
The initial field in these situations can be conceptualized as small-scale (on the order of a few electron or ion skin depths) elongated magnetic flux ropes associated with current filaments~\citep{silva2003interpenetrating,huntington2015observation}.
The ultimate fate of these seed fields is unclear. 
On one hand, although they may be strong
(in the sense that magnetic pressure is comparable to the plasma pressure, i.e., the plasma $\beta \sim 1$), they might dissipate rather quickly due to their small scales. 
On the other hand, current filaments are prone to coalesce and thus gradually transfer energy to larger scales.
Therefore, it is not {\it a priori} clear whether such electron-scale (or ion-scale) fields survive on long time-scales, how quick and effective the inverse magnetic-energy cascade arising from their mergers might be, and what fraction of the initial magnetic energy the inverse cascade can deliver to large-enough scales, where the field can be picked up and amplified by the ambient turbulence (e.g., in the context of Galactic magnetogenesis). 
An estimation of such scales can be found in Appendix~\ref{sec:scale_estimation}.
In the context of GRB prompt emission and afterglow, an important additional question is whether the Weibel-produced field in a relativistic shock can survive long enough to explain the observed powerful synchrotron emission~\citep{medvedev1999generation,silva2003interpenetrating,medvedev2004long,ruyer2018}. 

A complex, volume-filling network of magnetic flux ropes is also relevant for various heliospheric environments. In the solar corona, flux-tube dynamics is believed to be a key element in determining the complex magnetic structures that are observed~\citep{parker1983}, and in addressing longstanding problems such as coronal heating~\citep{velli1999,dmitruk1999,klimchuk2008,khabarova2015,zank2015}. 
In the solar wind, in-situ measurements made by the ACE, Ulysses and Wind Spacecraft track the evolution and merging of flux tubes, which may be associated with enhancement in the flux of energetic ions~\citep{hu2019}.
In addition, observations from the two Voyager spacecraft~\citep{stone2005voyager,stone2008asymmetric} and IBEX~\citep{mccomas2009global} strongly suggest that the heliosheath is composed of a turbulent sea of magnetic flux ropes~\citep{opher2011magnetic}.
Understanding the dynamics of these flux ropes is thus critical to developing a model of the heliosheath, as well as its efficiency at accelerating energetic particles. Indeed, magnetic reconnection in this ``sea'' of flux tubes has been proposed as a mechanism for the acceleration of anomalous cosmic  rays ~\citep{lazarian2009model,drake2010magnetic,leroux2015}, providing additional motivation to understanding the merging of flux tubes in a statistical way.

The broad recognition of the scientific importance of flux-rope dynamics has motivated numerous theoretical and experimental studies. 
Detailed numerical kinetic investigations of merging of a small number (usually 2-4) of flux ropes were carried out recently~\citep{east2015,yuan2016,lyutikov2017particle,lyutikov_sironi_komissarov_porth_2017}. 
These studies consider only a few interacting flux ropes and thus cannot be viewed as statistical investigations of a hierarchical, multi-stage coalescence process. 
However, in recent years, several theoretical studies have used statistical approaches to understand systems of merging structures in various astrophysical contexts~\citep{gruzinov2001gamma,medvedev2004long,kato2005saturation,katz2007self,fermo2010statistical,lyutikov2017particle,lyutikov_sironi_komissarov_porth_2017,zrake2017turbulent} and dynamics of coronal magnetic loops~\citep{uzdensky2008statistical}.
Despite all these efforts, important questions remain to be answered. 
For example, what are the underlying physical mechanisms of the merging process, and what are the merging statistics in 3D systems?
On the experimental side, there have been detailed investigations of the 2D dynamics during the merger of pairs of annular flux ropes~\citep{yamada1997study,jara2016laboratory} and others focusing on the 3D dynamics of flux ropes~\citep{furno2005coalescence,gekelman2012three,intrator2013flux,sears2014flux,gekelman2016pulsating}.

Another problem which may be intimately connected to the flux-tube dynamics is that of the recently reported numerical observation of inverse transfer of magnetic energy in non-helical turbulent systems~\citep{zrake2014inverse,brandenburg2015nonhelical,Reppin2017}.
These results have been met with some surprise, because conventionally in MHD turbulence, the inverse energy transfer associated with the inverse cascade of magnetic helicity is explained by the conservation of non-zero net magnetic helicity~\citep{pouquet_1978,matthaeus1986,christensson2001}.
A possible approach to understanding the above-mentioned recent numerical observations is to figure out the detailed magnetic field dynamics that enables inverse energy transfer. From this point of view, a statistical description of the dynamics of magnetic structures can shed light on this self-organization of magnetic field by providing a dynamical model of this process.  

Motivated by these considerations, in previous work~\citep[][hereafter referred to as Z19]{zhou2019magnetic} 
we derived a solvable analytical model describing the evolution of two-dimensional, initially small-scale magnetic fields via their successive coalescence enabled by magnetic reconnection.
The model is based on the conservation of mass and magnetic flux during the merging process.
Our theory identifies magnetic reconnection as the key mechanism controlling the evolution of the field and determining the properties of such evolution: magnetic energy is found to decay as $\tilde t^{-1}$, where $\tilde t$ is time normalized to the (appropriately defined) reconnection timescale; and the correlation length of the field grows as $\tilde t^{1/2}$. 
Our 2D hierarchical model was directly verified by 2D reduced-magnetohydrodynamics (RMHD) simulations that we carried out.

In the present paper, we generalize our theory to 3D systems (Section~\ref{sec:theory} and Appendix~\ref{sec:kink}), which we successfully test with 3D RMHD simulations (Section~\ref{sec:numerics}).
In Appendix~\ref{sec:scale_estimation}, we apply the scaling laws of magnetic field evolution obtained in this work to the problem of galactic magnetogenesis, and address the relevance of small-scale magnetic seed fields to the galactic dynamo.
Additionally, we demonstrate (Appendix~\ref{sec:noise}) that reconnection-enabled inverse transfer remains a robust phenomenon when the initial condition is a random magnetic field, rather than a structured array of magnetic flux tubes.
Conclusions from this study and a discussion of some implications of our work can be found in Section~\ref{sec:conclusions}.

\section{Statistical theory of three-dimensional flux-tube mergers}
\label{sec:theory}
\subsection{Three-dimensional hierarchical model}
\label{sec:3d-model}
In this section, we introduce a 3D hierarchical model to describe the merger of magnetic flux tubes. 
Our goal is to obtain explicit expressions for the evolution of quantities such as magnetic energy and the coherence length of magnetic field in both the merging plane and the direction perpendicular to it, as functions of time.
Similar to its 2D counterpart~\citepalias{zhou2019magnetic}, the 3D model is constructed in the framework of resistive incompressible RMHD, although the key ideas should remain valid in some collisionless plasma descriptions.

\begin{figure}
    \centering
    \includegraphics[width=0.6\textwidth]{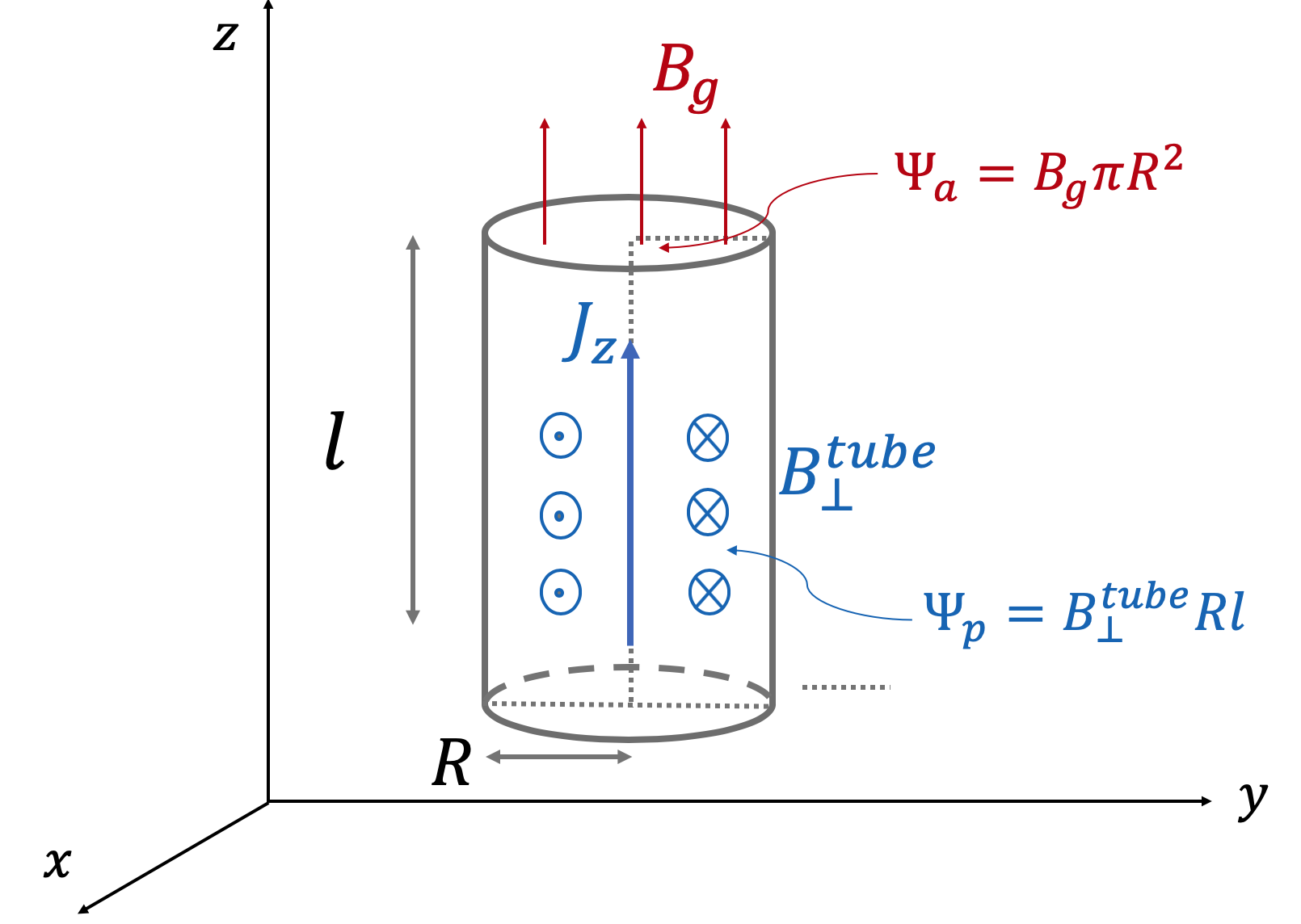}
    \caption{Schematic of the flux tube geometry considered in our analytical model. A flux tube is characterized by its radius $R$, length $l$, perpendicular magnetic field $B_\perp^{\rm tube}$, axial flux $\Psi_a$ and poloidal flux $\Psi_p$.}
    \label{fig:schem1}
\end{figure}

In our RMHD system, a constant strong background magnetic guide field $\mathbf{B}_g=B_g \hat{\mathbf{z}}$ is assumed, and the $xy$ plane is referred to as the perpendicular plane.  
We consider an ensemble of magnetic flux tubes of the same size and with the same magnetic field strength as the initial condition.
The magnetic field in each flux tube can be decomposed into the constant guide field $\mathbf{B}_g$ and the perpendicular magnetic field  $\mathbf{B}_\perp$, corresponding to an electric current parallel or anti-parallel to the guide field. For simplicity, each flux tube is assumed to be of cylindrical shape, with radius $R$ (in the $xy$ plane) and length $l$ in the $z$ direction.
The lengths $R$ and $l$ represent the coherence lengths of the perpendicular magnetic field, $\mathbf{B}_\perp$, in the directions perpendicular and parallel, respectively, to the guide field.
In the RMHD approximation, $B_\perp$ and $R$ are asymptotically smaller than $B_g$ and $l$ respectively.
The magnetic field lines in each flux tube thus have either right- or left-handed helical shapes depending on the direction of the current, with asymptotically small pitch angles. The flux tubes are assumed to be randomly distributed in the system and volume-filling, with half of them carrying currents  parallel to the guide field, and the other half anti-parallel.
Each flux tube is a system of concentric cylindrical magnetic flux surfaces nested around the tube's single magnetic axis aligned with the guide field.
It contains axial magnetic flux $\Psi_a=B_g\pi R^2$ and poloidal (i.e., azimuthal around the magnetic axis) magnetic flux $\Psi_p=B_\perp^{\rm tube} Rl$, where $B_\perp^{\rm tube}$ is the characteristic perpendicular magnetic field of each flux tube. 
In the perpendicular $xy$ plane, the magnetic potential of a flux tube is defined as $\psi^{\rm tube}=B_\perp^{\rm tube} R$, similar to its 2D counterpart. Thus, $\Psi_p=\psi^{\rm tube} l$.
Fig.~\ref{fig:schem1} schematically represents the simplified geometry of a single tube as we have just described.

The magnetic energy associated with the perpendicular magnetic field contained in a flux tube is $\epsilon_M \simeq \pi R^2l(B_\perp^{\rm tube})^2/(8\pi)=(\psi^{\rm tube})^2 l/8$.
Denoting by $N$ the number density of flux tubes per unit volume [$N \sim 1/(\pi R^2l)$], the total magnetic energy density (of the perpendicular magnetic field) of the system is, therefore, $E_M=\epsilon_M N = (B_\perp^{\rm tube})^2/(8\pi)$.
In addition, we denote the number density of flux tubes per unit area in the $xy$ plane as $N_{\perp}$, where $N_\perp \sim 1/(\pi R^2)$ and $N_\perp = Nl$.

A non-zero local magnetic helicity exists associated with each flux tube, defined as $h= \int dV_{\rm tube} \mathbf{A} \cdot \mathbf{B}$, where $\mathbf{B}=\mathbf{\nabla}\times\mathbf{A}$. 
In the RMHD limit, however, this expression simplifies as follows:
\begin{equation}
\begin{aligned}
\label{eq:rmhd_helicity}
    h &= \int dV_{\rm tube} \mathbf{A} \cdot \mathbf{B} = \int dV_{\rm tube} (A_z B_g + A_\perp B_\perp^{\rm tube}) \\
    &\approx -B_g \int dV_{\rm tube} \psi^{\rm tube} \approx -(B_g \pi R^2)(B_\perp^{\rm tube} Rl) = -\Psi_a \Psi_p, 
\end{aligned}
\end{equation}
represented by the product of axial flux and poloidal flux. In the derivation, the approximation is made based on the RMHD ordering $A_\perp \ll A_z$ and $B_\perp^{\rm tube} \ll B_g$.

When two flux tubes with parallel axial currents are close to each other, they are unstable to the coalescence instability and approach each other~\citep{finn1977coalescence}, merging into a wider flux tube via reconnection of their poloidal fluxes.
For simplicity, we assume the merging process to happen in discrete stages, denoted by index $n$. 
We assume the mergers to proceed in hierarchical fashion and, therefore, take all flux tubes to be identical at each stage. 
The index $n$ is thus added as a subscript to the above quantities to indicate the generation of mergers to which they belong.
As the flux tubes merge, the above quantities evolve; i.e., they are functions of $n$ and so, implicitly, of time. 

We also assume mergers to be binary ($N_{\perp,n+1}=N_{\perp,n}/2$) and governed by basic conservation laws [similar arguments can be found in \citet{medvedev2004long,polomarov2008,lyutikov2017particle,zrake2017turbulent}].

First, the merger of two flux tubes should conserve the axial magnetic flux: $\Psi_{a,n+1}=2\Psi_{a,n}$. Since the axial magnetic field $\mathbf{B}_g$ is a constant, it follows that the area of the perpendicular cross section should be conserved and, therefore,
\begin{equation}
    \label{eq:radius}
    R_{n+1} = \sqrt{2} R_n.
\end{equation} 

Second, the perpendicular magnetic potential associated with each flux tube is not affected by reconnection and, as in the two-dimensional case~\citepalias{zhou2019magnetic}, should remain the same: $\psi_{n+1}^{\rm tube}=\psi_n^{\rm tube}$.
Combined with Eq.~(\ref{eq:radius}), we obtain the evolution of perpendicular magnetic field and its energy density:
\begin{equation}
    \label{eq:B_recursion}
    B_{\perp,n+1}^{\rm tube}=B_{\perp,n}^{\rm tube}/\sqrt{2}, \qquad E_{M,n+1} = E_{M,n}/2.
\end{equation}

We assume that the parallel coherence length of each flux tube is determined by the critical balance condition, i.e., the statement that the pertinent parallel and perpendicular timescales should roughly be the same~\citep{Goldreich1995}.
Parallel dynamics is dictated by the propagation of shear Alfv\'en waves along the magnetic field. Therefore, the parallel timescale is $\tau_\parallel\sim l/V_{A}$, with $V_A$ the Alfv\'en speed computed with the guide field.
In the perpendicular plane, the relevant timescale is $\tau_\perp\sim R/v_A$, with $v_A$ the Alfv\'en speed computed with $B_\perp^{\rm tube}$.
That this is the relevant perpendicular timescale to consider remains true even during mergers, since merging tubes are not static, but rather move in the perpendicular plane at roughly the ambient $v_A$. 
Note also that another relevant perpendicular timescale is the reconnection time --- that is, the time it takes to merge two tubes through reconnection. This is always slower than Alfv\'enic, implying that critical balance can always establish itself dynamically.

We thus find
\begin{align}
\label{eq:critical_balance}
   \tau_{\parallel}\sim \tau_\perp \Rightarrow l/B_g \sim R/B_\perp^{\rm tube}.
\end{align}
Plugging Eqs.~(\ref{eq:radius}-\ref{eq:B_recursion}) into this relation, we obtain the evolution of coherence length of magnetic field in the parallel direction:
\begin{equation}
    \label{eq:parallel_length}
    l_{n+1}=2l_n.
\end{equation}
That is, $l \sim R^2$, i.e., the magnetic flux tubes grow faster in the parallel direction than on perpendicular planes. 
Knowing the magnetic potential remains the same during the merging process, the poloidal flux ($\Psi_p=\psi^{\rm tube}l$) increases with the length of the flux tubes: $\Psi_{p,n+1}=2\Psi_{p,n}$. As a result of the increasing flux contained in each generation of flux tubes as they merge and grow, the magnetic helicity associated with each flux tube also increases: $h_{n+1}=4h_n$.

Our next step in order to obtain a continuous-time description of the evolution is to compute the lifetime of each generation; this entails a discussion of the reconnection process mediating the mergers.

The Lundquist number corresponding to the merging tubes at any given $n$-th stage is defined as $S_n \equiv R_n v_{A,n}/\eta$, where $\eta$ is the (constant) magnetic diffusivity. 
Importantly, $S \propto RB_\perp^{\rm tube}/\eta \propto \psi^{\rm tube}/\eta $. 
Therefore, and as we pointed out in our earlier work~\citepalias{zhou2019magnetic}, the constancy of the magnetic potential $\psi^{\rm tube}$ during the mergers implies the same for the Lundquist number: $S_{n+1}=S_n$.
In particular, in MHD, if $S \lesssim 10^4$, reconnection proceeds in the Sweet-Parker (SP) regime~\citep{sweet_neutral_1958,parker_sweet_1957}.
In this regime, the dimensionless reconnection rate, defined as $\beta_{\rm rec} \equiv v_{\rm rec}/v_A$, scales with the Lundquist number as $\beta_{\rm rec} \simeq S^{-1/2}$.
In contrast, if $S \gtrsim 10^4$, then magnetic reconnection proceeds in the plasmoid-dominated regime~\citep{biskamp-1986,loureiro2007instability,lapenta2008self,samtaney2009formation,bhattacharjee2009fast,uzdensky2010fast,huang2010scaling,loureiro2012magnetic,loureiro_magnetic_2016} with $\beta_{{\rm rec}}\simeq 0.01$.
Therefore, we reach the non-trivial conclusion that the normalized reconnection rate, $\beta_{\rm rec}$, remains unchanged throughout the evolution, and the merging process remains in the same reconnection regime (SP or plasmoid-dominated) in which it started initially.

We can now proceed to calculate the lifetime of each generation.
We note first that, by definition, the time it takes to reconnect (merge) two flux tubes is $\tau_{\rm rec} = \beta_{\rm rec}^{-1} R/v_A=\beta_{\rm rec}^{-1}\tau_\perp$.
Since $\beta_{\rm rec} \ll 1$, the time scale for the $n$-th stage is approximately the reconnection time at this stage:
\begin{align}
    \tau_{n} \approx \tau_{\rm rec,n } \propto  \beta_{\rm rec}^{-1} R_{n}/B_{\perp,n}^{\rm tube}.
\end{align}

Consequently, $\tau_{n+1} = 2\tau_n$.
The time taken to reach the $n$-th stage is thus:
\begin{align}
&t_{n} = \sum_{k=0}^{n-1} \tau_k = \tau_0 \sum_{k=0}^{n-1} 2^{k} \approx \tau_0 2^{n}, \quad n\gg 1, \label{eq:time}
\end{align} 
where $\tau_0 =  \beta_{\rm rec}^{-1} R_{0}/B_{\perp,0}^{\rm tube}$, $R_0$ and $B_{\perp,0}^{\rm tube}$ are the radius and perpendicular magnetic field of the initial flux tubes respectively.  
Therefore we can express the index $n$ as a function of time: $n = 
\log_2 t/\tau_0$.
The explicit formulae for the different quantities as a function of $n$ can be obtained from the recursive relations derived earlier:
\begin{align}
& B_{\perp,n}^{\rm tube} = B_{\perp,0}^{\rm tube} 2^{-n/2}, \quad E_{M,n} = E_{M,0}  2^{-n}, \\
&k_{\perp,n} = k_{\perp,0} 2^{-n/2}, \quad R_n = R_0 2^{n/2}, \quad l_n = l_0 2^{n}, \\
& N_{\perp,n} = N_{\perp,0} 2^{-n}.
\end{align}
By using $n = \log_2 t/\tau_0$, we obtain the continuous time evolution of the relevant quantities:
\begin{align}
& B_\perp^{\rm tube} = B_{\perp,0}^{\rm tube} \tilde{t}^{-1/2}, \quad E_M = E_{M,0}  \tilde{t}^{-1}, \label{eq:field_scalings}\\
&k_\perp = k_{\perp,0} \tilde{t}^{-1/2}, \quad R = R_0 \tilde{t}^{1/2}, \quad l = l_0 \tilde{t}, \label{eq:length_scalings}\\
& N_{\perp} = N_{\perp,0} \tilde{t}^{-1},\label{eq:number_scalings}
\end{align}
where $\tilde{t} \equiv t/ \tau_0$ is time normalized to the initial reconnection timescale and $k_\perp \equiv 2\pi/R$ is the perpendicular wavenumber.
We note that the aspect ratio of a flux tube $l/R \sim \tilde{t}^{1/2}$. Therefore, as flux tubes merge and grow, they become progressively more slender. However, they always satisfy the critical balance and remain (marginally) stable to kink-type instabilities (Appendix~\ref{sec:kink}). 
We also note that $l = l_0 \tilde{t} = l_0 (v_{A,0} \beta_{\rm rec}/R_0) t$.
Combined with the critical balance scaling of initial flux tubes $l_0/V_A \sim R_0/ v_{A,0}$, where $V_A$ is the Alfv\'en speed based on the guide field, we obtain $l  = V_A \beta_{\rm rec} t$. We can thus consider $V_A \beta_{\rm rec}$ as the characteristic speed for the growth of tube's length. 

To summarize our 3D hierarchical model: in the perpendicular plane, the merging of flux tubes through reconnection results in the decay of the perpendicular magnetic field, and in the growth of the perpendicular length scale [Eq.\eqref{eq:field_scalings} and Eq.\eqref{eq:length_scalings}]; these conclusions are identical to those obtained from two-dimensional considerations [namely, Eq.(3) in~\citetalias{zhou2019magnetic}]. 
In the parallel direction (i.e., along the guide field), the dynamics are (causally) set by critical balance; combined with the perpendicular relations, this yields a prediction for the growth of the flux tube coherence length in the direction along the guide field.

\subsection{Self-similar properties and magnetic power spectra}
\label{sec:selfsimilar-scaling}
The power-law time dependencies [Eqs.~\eqref{eq:field_scalings}--\eqref{eq:number_scalings}] suggest that the system evolves in a self-similar manner. 
We note that the perpendicular dynamics in the 3D RMHD system is the same as that in its 2D counterpart~\citepalias{zhou2019magnetic}. Therefore, the self-similar properties and the behavior of perpendicular magnetic power spectrum should also be identical between the 2D and 3D cases.
We discuss these properties in more detail in this section, for the completeness of this paper.
In the perpendicular plane, the growing length scale and decreasing field strength [Eq.\eqref{eq:field_scalings} and Eq.\eqref{eq:length_scalings}] can be represented by a dynamical renormalization, which we show as follows. 
With the power-law time evolution, quantities at any given two moments of time $\tilde{t}$ and $\tilde{t'}$ can be related through a scaling factor $\lambda=(\tilde{t'}/\tilde{t})^{1/2}$
\begin{align}
    k_\perp(\tilde{t}) = \lambda k_\perp(\tilde{t'}), \quad B_\perp^{\rm tube}(\tilde{t})= \lambda B_\perp^{\rm tube}(\tilde{t'}), \quad \tilde{t}=\lambda^{-2} \tilde{t'}.
\label{eq:scale_transform1}
\end{align}
That is, the time evolution of the system is equivalent to performing the above scale transformation. 
In a system consisting of volume-filling flux tubes, the perpendicular magnetic field $\mathbf{B}_\perp$ follows the same scaling as $B_\perp^{\rm tube}$. That is, $\mathbf{B}_\perp(\tilde{t})= \lambda \mathbf{B}_\perp(\tilde{t'})$.

Let us see how this self-similarity affects the evolution of perpendicular magnetic power spectrum, which we define as
\begin{equation}
    M(k_\perp,\tilde{t}) \equiv  \frac{1}{8 \pi}\frac{2 \pi k_\perp}{(2 \pi)^2} \int d^2 \mathbf{r}
    e^{i \mathbf{k_\perp} \cdot\mathbf{r}} \langle \mathbf{B}_\perp(\mathbf{x},\tilde{t}) \cdot \mathbf{B}_\perp(\mathbf{x}+\mathbf{r},\tilde{t}) \rangle ,
    \label{eq:defineU}
\end{equation}
where the angle brackets represent a 3D average over all $\mathbf{x}$, and $\mathbf{r}$ is a vector on the perpendicular ($x$-$y$) plane. The spectrum is also averaged over the $z$ dimension, which is not explicitly shown in Eq.~\eqref{eq:defineU}.
Using Eq.~\eqref{eq:scale_transform1}, the spectra at different times are related as
\begin{equation}
\begin{aligned}
\label{eq:rescalingM}
&M[k_\perp(t'),\tilde{t'}]=M(k_\perp/\lambda, \lambda^2 \tilde{t}) = \frac{1}{8 \pi}\frac{2 \pi (k_\perp/\lambda)}{(2 \pi)^2} \int d^2 (\lambda \mathbf{r})\\
    &e^{i (\mathbf{k_\perp}/\lambda) \cdot (\lambda \mathbf{r})} \langle \mathbf{B}_\perp(\lambda \mathbf{x},\lambda^2 \tilde{t}) \cdot \mathbf{B}_\perp[\lambda (\mathbf{x}+\mathbf{r}),\lambda^2 \tilde{t}] \rangle \\
    &= \lambda^{-1}\frac{1}{8 \pi}\frac{2 \pi k_\perp}{(2 \pi)^2} \int d^2 \mathbf{r} 
    e^{i \mathbf{k_\perp} \cdot \mathbf{r}} \langle \mathbf{B}_\perp(\mathbf{x},\tilde{t}) \cdot \mathbf{B}_\perp(\mathbf{x}+\mathbf{r},\tilde{t}) \rangle \\
    &=\lambda^{-1} M(k_\perp,\tilde{t}).
\end{aligned}
\end{equation}
It can be shown that the perpendicular magnetic power spectrum satisfying the relation Eq.~\eqref{eq:rescalingM} has the self-similar solution~\citep{olesen1997inverse}
\begin{equation}
\begin{aligned}
    \label{eq:U_scalingfunction}
    M(k_\perp,\tilde{t})&=\tilde{t}^{-1/2}\bar{M}(k_\perp\tilde{t}^{1/2}) 
    =k_\perp F(k_\perp\tilde{t}^{1/2}),
\end{aligned}
\end{equation}
where $\bar{M}$ and $F$ are scaling functions of the variable $k_\perp \tilde{t}^{1/2}$. 

The solution Eq.~\eqref{eq:U_scalingfunction} represents the pattern of energy decay and scale growth, determined by the physics of the merger events. However, it does not contain any information about energy as a function of scale. 
Assuming the case of a power-law spectrum, the scaling function has to satisfy $\Bar{M}(k_\perp \tilde{t}^{1/2}) \propto (k_\perp\tilde{t}^{1/2})^{-\gamma}$ where $-\gamma$ is the index of the spectrum. Eq.~\eqref{eq:U_scalingfunction} can thus be further expressed as 
\begin{equation}
\begin{aligned}
    \label{eq:U_k_t}
    M(k_\perp,\tilde{t})&\propto {t}^{-1/2} (k_\perp\tilde{t}^{1/2})^{-\gamma} 
     \propto \tilde{t}^{-\alpha} k_\perp^{-\gamma},
\end{aligned}
\end{equation}
where $2\alpha=\gamma+1$.

\section{Numerical study}
\label{sec:numerics}
We now proceed to test the predictions of the previous section by means of direct numerical simulations with the code {\tt Viriato}~\citep{loureiro2016viriato}.
The code solves the three-dimensional, incompressible, reduced-MHD equations~\citep{kadomtsev1973nonlinear,strauss1976nonlinear,zank1992,schekochihin2009astrophysical}:
\begin{align}
\label{eq:induction}
    &\frac{\partial \psi}{\partial t} + [\phi, \psi] = V_A \frac{\partial \phi}{\partial z}+  \eta \nabla_\perp^2 \psi, \\
\label{eq:momentum}    
    &\frac{\partial}{\partial t}  \nabla_\perp^2 \phi + [\phi,\nabla_\perp^2 \phi]= V_A \frac{\partial}{\partial z}\nabla^2_\perp \psi + [\psi, \nabla_\perp^2 \psi] + \nu \nabla_\perp^4 \phi,
\end{align}
where $V_A$ is the Alfv\'en speed based on the (constant and uniform) guide-field, $V_A=B_g/\sqrt{4 \pi \rho}$, and  $[A,B] \equiv \partial_x A \partial_y B - \partial_y A \partial_x B$.
The flux function $\psi$ is defined as $\psi \equiv -A_z/\sqrt{4\pi \rho}$, where $A_z$ is the axial component of the magnetic vector potential.
The stream function  $\phi$ is $\phi \equiv c\varphi/B_g$ where $\varphi$ is the electrostatic potential. 
The magnetic field (measured in velocity units) and velocity are thus related to $\psi$ and $\phi$ as:
\begin{equation}
\begin{aligned}
\label{eq:bu_psi_phi}
    &\mathbf{B}= \hat{\mathbf{z}} \times \mathbf{\nabla} \psi + \hat{\mathbf{z}} B_g \\
    &\mathbf{u}_\perp= \hat{\mathbf{z}} \times \mathbf{\nabla} \phi.
\end{aligned}
\end{equation}

The viscosity $\nu$ is taken equal to the magnetic diffusivity $\eta$ in all simulations.
\subsection{Simulation setup}
In {\tt Viriato}, the length scales on the perpendicular plane and in the parallel direction are normalized to (arbitrary) reference lengths $L_\perp$ and $L_\parallel$ respectively ($L_\perp \ll L_\parallel$ in the RMHD limit). 
Times are normalized to the parallel Alfv\'en time $\tau_A=L_\parallel/V_A$.
In the following description, all quantities are given in dimensionless form.
The simulation domain is a box of dimensions $L_x \times L_y \times L_z$, where $L_x=L_y$ in all the simulations presented. 
 We set $L_x=2\pi$, and show the results from simulations with $L_z=4L_x$. 
A more elongated simulation box allows more generations of mergers before the length $l$ of flux tubes in the $z$ direction reaches the length of the box. In order to adequately resolve the dynamics along $z$, and considering the computational cost of the simulation, we find the box with $L_z=4L_x$ to be optimal. 
The box has periodic boundary conditions in all directions. 

The initial condition is a straightforward extension to 3D of our previous 2D study~\citepalias{zhou2019magnetic}.
There is no initial flow, $\phi(x,y,z, t=0)=0$, and the magnetic flux tubes are invariant in the $z$ direction: $\psi(x,y,z,t=0) = \psi_0 \cos( k_0 x) \cos(k_0 y)$.
It is easy to check that this initial condition satisfies the dissipationless version of Eqs.~(\ref{eq:induction}-\ref{eq:momentum}) and is therefore an (ideal) equilibrium. 
We thus have a $2k_0\times 2k_0$ static array of guide-field-aligned magnetic flux tubes of alternating polarities. In all runs we set $k_0=8$ and so the initial radius of each flux tube is $R_0=L_x/4k_0=\pi/16$. 
We further fix $\psi_0 k_0=1$ to set the strength of the initial in-plane magnetic field $B_{\perp,0}\equiv\psi_0/R_0=2/\pi$.
A spatially-random perturbation with small (linear) amplitude is added to trigger the dynamic evolution of the system. This perturbation breaks the axial symmetry of the system.

In the RMHD limit, the magnetic helicity of the system is $H_M=\int dV \mathbf{A} \cdot \mathbf{B} \approx B_g \int dV \psi $. 
We note that while the local magnetic helicity associated with each flux tube is finite [Eq.~\eqref{eq:rmhd_helicity}], the net magnetic helicity of the entire setup is zero.
Since $H_M$ is an ideal invariant, it should remain approximately zero throughout the evolution of the system. 
Therefore, our simulations are relevant to the study of inverse energy transfer in non-helical turbulent systems, a question briefly alluded to in the Introduction and further discussed in Appendix~\ref{sec:noise}.

We set $\nu=\eta=10^{-4}$ in the simulation, corresponding to the Lundquist number of the initial flux tubes $S_0\equiv R_0 v_{A,0}/\eta=1250$. 
The perpendicular plane is (spectrally) resolved with $1024^2$ grid points.
As per our discussion of Section~\ref{sec:3d-model}, this value of $S_0$ implies that we expect the reconnection of two flux tubes to fall in the Sweet-Parker (SP) regime; the thickness of the SP current layers during the first generation of mergers, $\delta_{SP,0}\approx S_0^{-1/2}R_0$ is only marginally resolved with about one cell.
But we note that as the system evolves and the flux tubes grow to larger scales, the thickness of the current sheets increases (the Lundquist number is expected to remain constant, as we have argued in Section~\ref{sec:3d-model}, but the radius of the tubes progressively increases) and the simulation becomes progressively better resolved.
Naturally, it would be desirable to investigate initial conditions characterized by even higher values of $S_0$, such as to access the plasmoid-mediated reconnection regime. 
Unfortunately, the computational requirements presented by such simulations are too great: we estimate that achieving $S_0  \gtrsim 10^4$ with the same number of initial flux tubes ($k_0=8$) requires a numerical resolution of at least 3000 grid points in each perpendicular dimension. This is too demanding even if the parallel resolution requirements were to remain unchanged.

The parallel length of the box is resolved with 512 grid points. A  seventh-order upwind scheme is used to advect the fields in this direction. 
To check the convergence with respect to parallel resolution, we performed a simulation with the same physical parameters but with $1024^3$ grid points instead (thus doubling the resolution in the $z$-direction), and evolved the system for the first few generations of mergers (during which the simulation is least resolved).
The results are indistinguishable from the fiducial run with $1024^2\times 512$ grid points.

Finally, we also perform a $512^3$ run in which the resistivity and viscosity are replaced with hyper-dissipation. 
Specifically, the dissipative terms in Eq.~\eqref{eq:induction} and Eq.~\eqref{eq:momentum} become $\eta_H \nabla_\perp^6 \psi$ and $\nu_H \nabla_\perp^{12} \phi$, respectively. Since the timestep with which the equations are integrated is constantly recomputed by {\tt Viriato} to maximize the efficiency of the calculation, it follows that the values of $\eta_H$ and $\nu_H$ are also continuously adjusted in the code  to ensure that the hyper-dissipation is reduced to the minimum possible for a given spatial resolution~\citep{loureiro2016viriato}. However, in the simulation reported here, we observe that the temporal variation of these coefficients is so small that it is justified to consider $\eta_H$ and $\nu_H$ as constants. The corresponding effective hyper-Lundquist number is $S_{H,0} \equiv R_0^5 v_{A,0}/\eta_H \approx 5 \times 10^8$.

\subsection{Visualizations}
\begin{figure}
\includegraphics[width=\textwidth,clip]{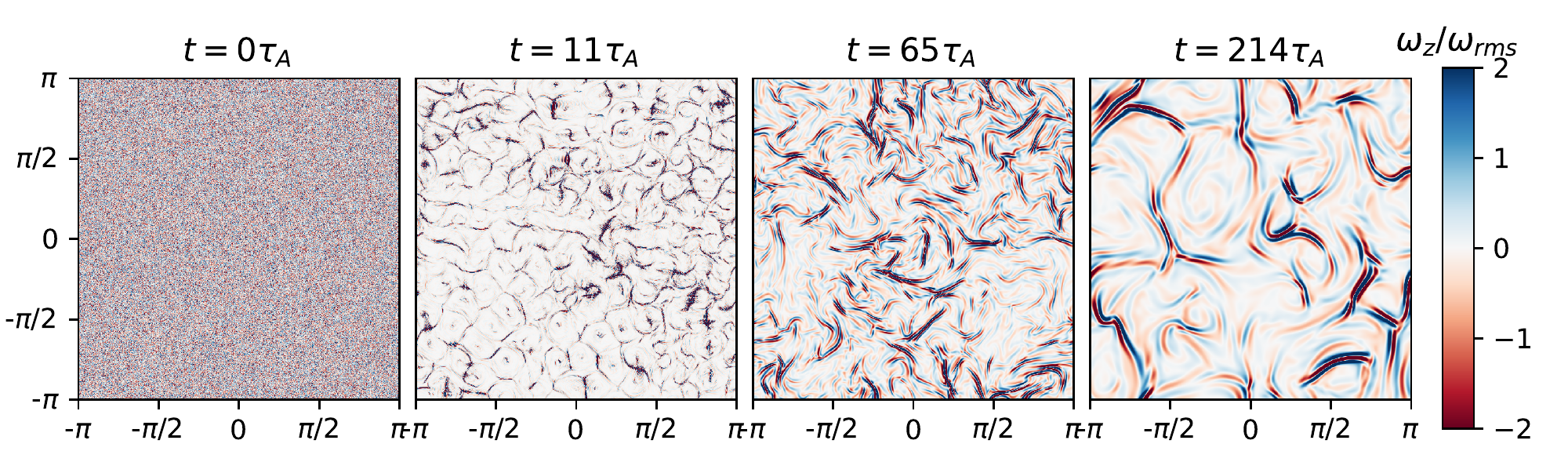}
\includegraphics[width=\textwidth,clip]{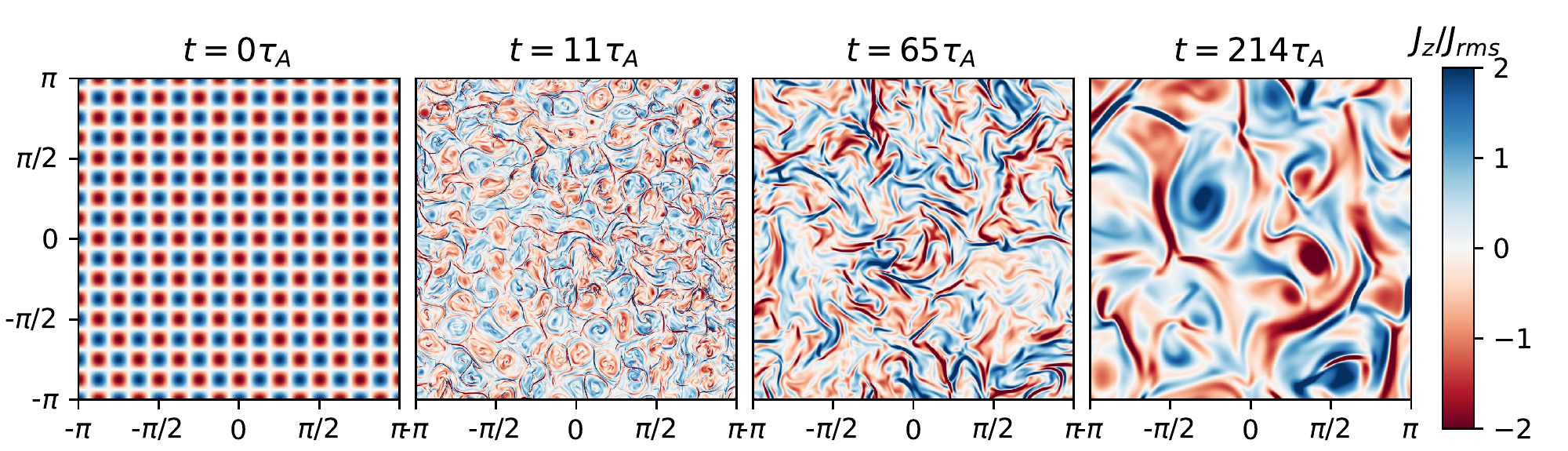}\\
\includegraphics[width=\textwidth,clip]{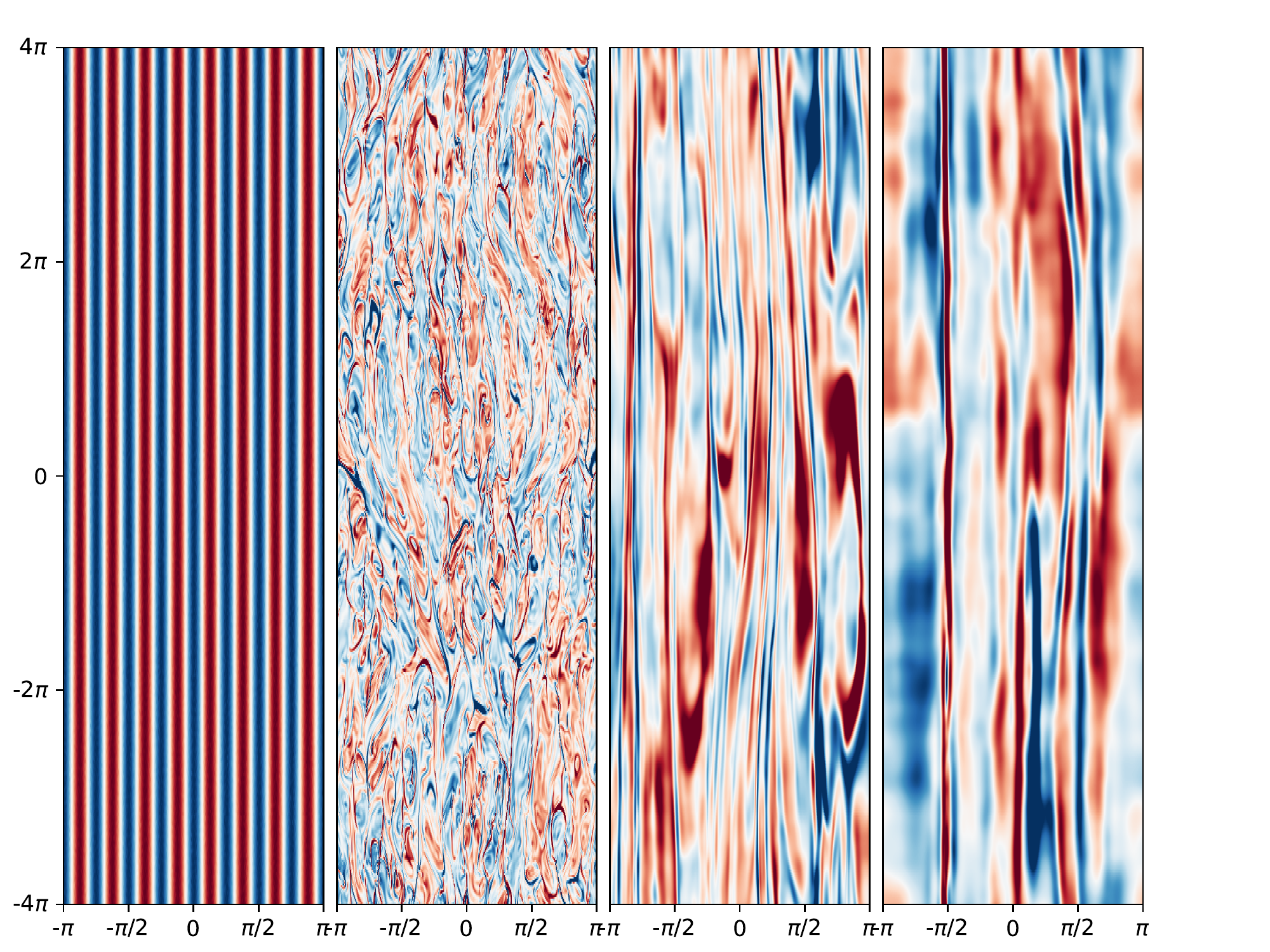}
\caption{Visualizations for the run with $S_0=1250$: vorticity density $\omega_z$ (normalized by its root mean square value) at various times on the $xy$ planes at $z=0$ (top); current density $J_z$ (normalized by its root mean square value) at various times on the $xy$ planes at $z=0$ (middle) and on the $xz$ planes at $y=0$ (bottom).}
\label{fig:2d_vis}
\end{figure}

We begin the analysis of the results of the numerical simulations by showing visualizations at different times of the run with regular dissipation, $S_0=1250$. 
The hyper-dissipative run looks qualitatively similar, except that structures are, of course, sharper. 
In Fig.~\ref{fig:2d_vis} we show the vorticity $\omega_z \equiv \nabla_\perp^2 \phi$ and current density $J_z \equiv \nabla_\perp^2 \psi$ at various times. 
The top panel shows $\omega_z$ on the $xy$ plane at $z=0$ [i.e., $\omega_z(x,y,z=0)$], and the middle panel shows $J_z$ on that same plane [i.e., $J_z(x,y,z=0)$].
The bottom panel shows $J_z$ on the $xz$ plane at $y=0$ [i.e., $J_z(x,y=0,z)$], plotted at the same times as the middle panels. 

\begin{figure}
\includegraphics[width=0.5\textwidth,clip]{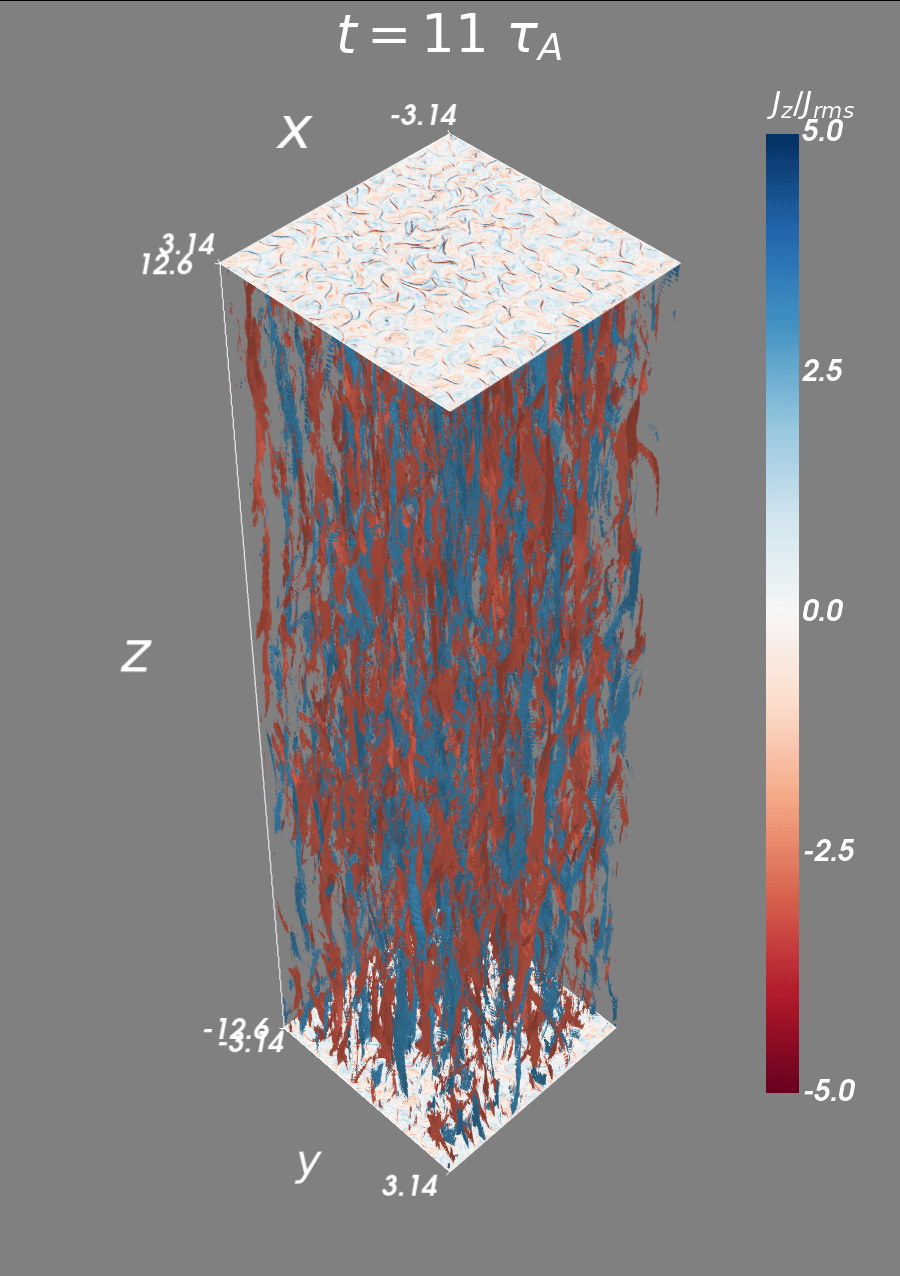}
\includegraphics[width=0.5\textwidth,clip]{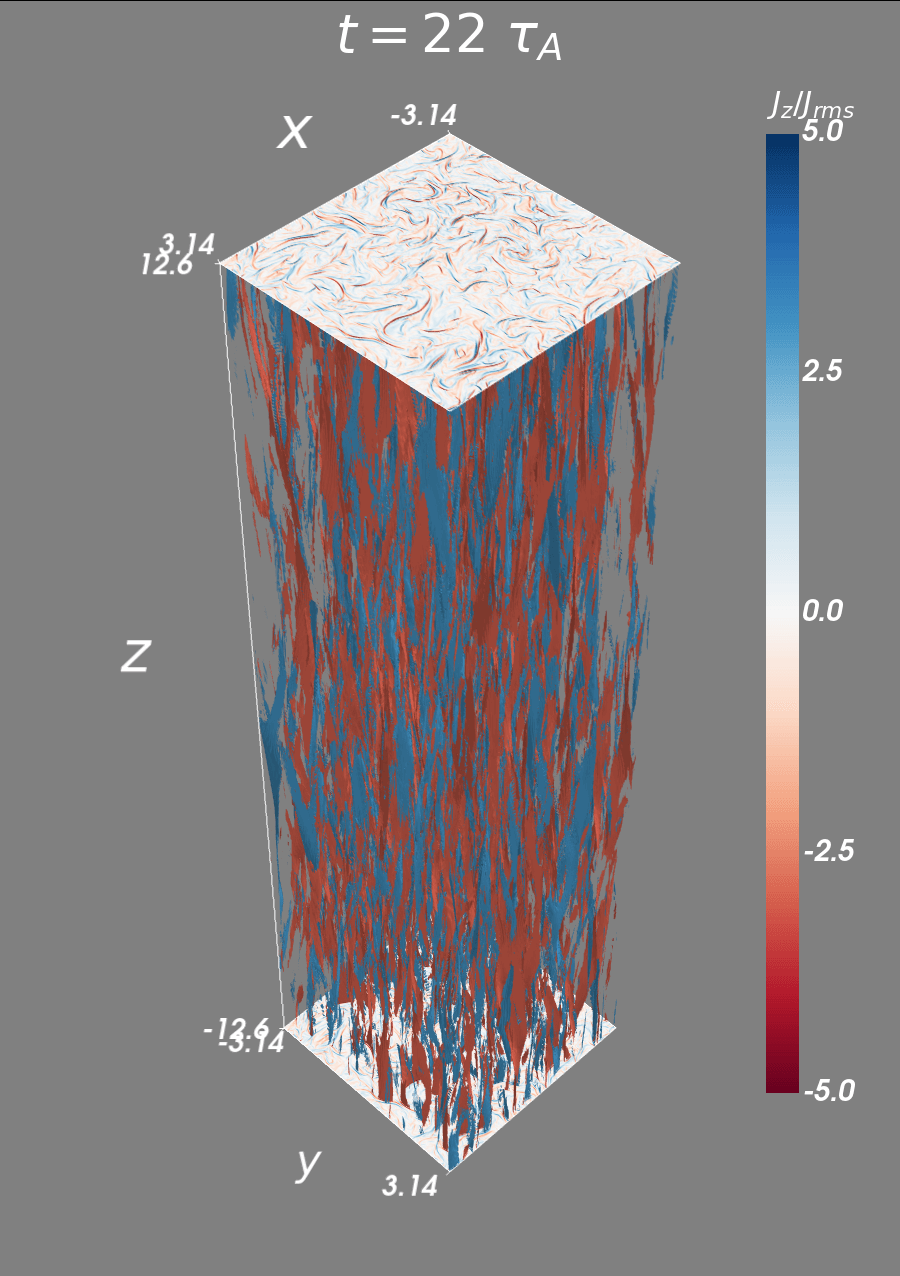}
\includegraphics[width=0.5\textwidth,clip]{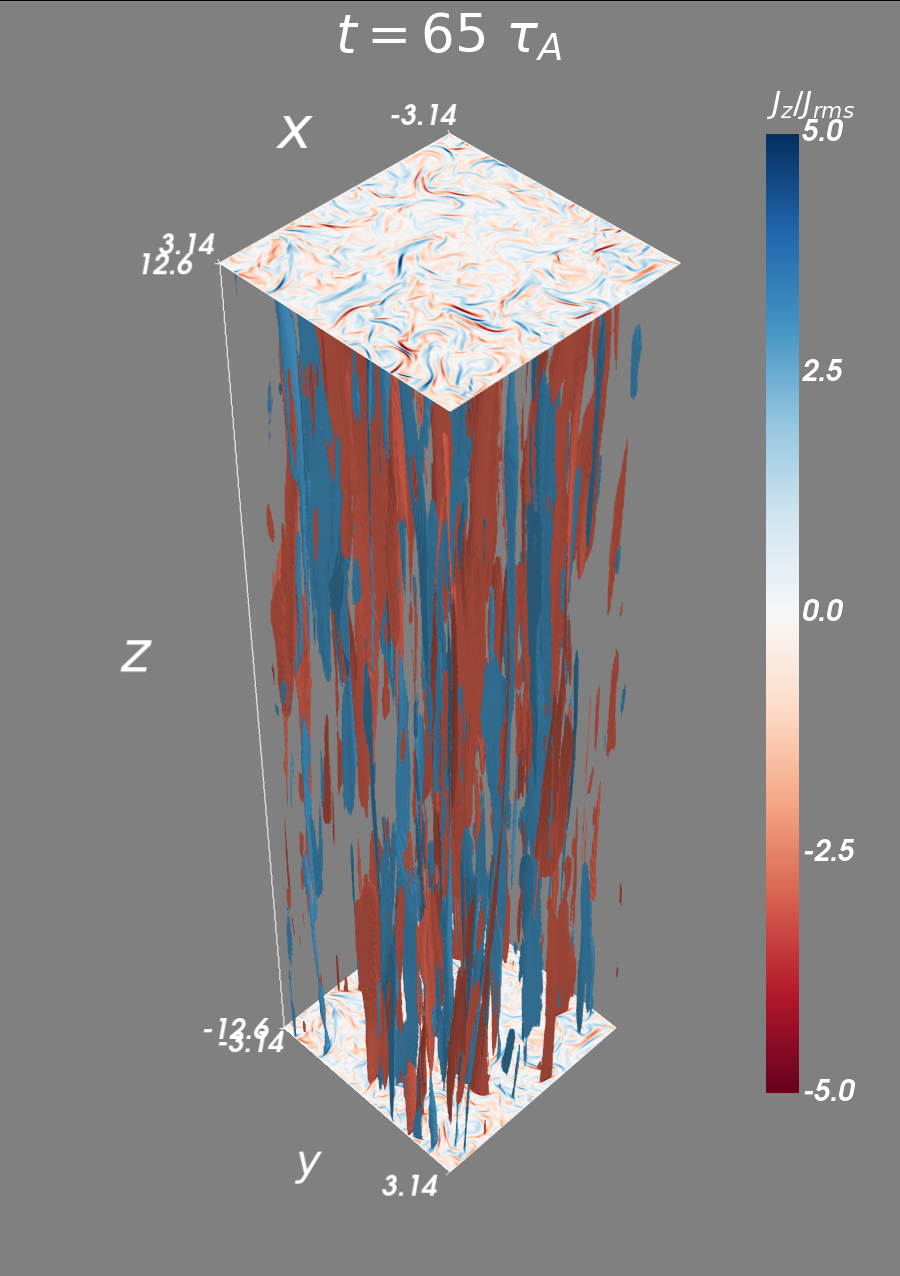}
\includegraphics[width=0.5\textwidth,clip]{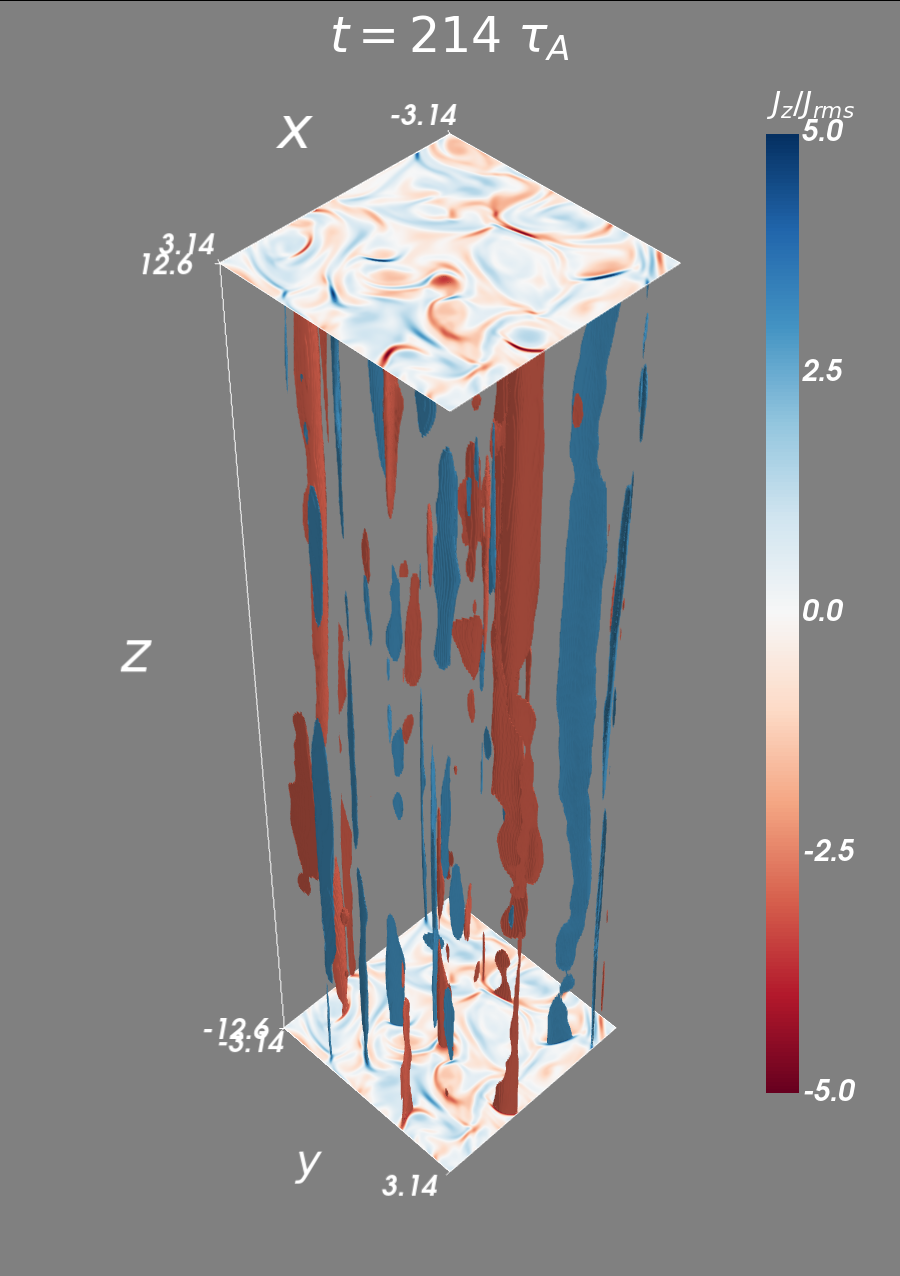}
\caption{Current density contours selected according to $|J_z|>3J_{\rm rms}$ at various times for the run with $S_0=1250$. }
\label{fig:curret_layers}
\end{figure}

These plots suggest a two-stage evolution of the system.
The length of the initial flux tubes ($8\pi$) is much longer than the length that critical balance would dictate (about $0.3$) and, therefore, the flux tubes are not causally connected from one end of the box to the other.
When reconnection events start at different positions along $z$-axis (i.e., on different $xy$ planes), the flux tubes break in the $z$-direction (kink-type instabilities are not expected in our RMHD system, as we discuss in Appendix~\ref{sec:kink}).
Due to this unstable set-up of elongated flux tubes, the system first enters a transient stage of developing turbulence.
At this stage, finer structures appear and the system becomes fully turbulent (see the snapshot at $t=11\tau_A$).
After the transient stage, the system relaxes to a more natural state, where the initial setup is ``washed out'' by the developing turbulence. 
The system then enters a prolonged stage of decaying turbulence, during which progressively larger flux tubes emerge (both in the $xy$ and in $xz$ planes). 
The length of the flux tubes grows faster than the radius, consistent with our prediction in Eq.~\eqref{eq:length_scalings} (This can be seen in Fig.~\ref{fig:2d_vis}, and a quantitative demonstration is shown in Section~\ref{sec:structurefunction}).
 
In Fig.~\ref{fig:curret_layers} we show current sheets, defined as the regions where $|J_z(x,y,z)|>3J_{\rm rms}$ and $J_{\rm rms}$ is the root mean square of $J_z$. The increasing complexity at the transient stage is manifested by this plot: the breaking of flux tubes leads to increasing number of current sheets, indicating more structures and reconnection sites.  
This phenomenon is also shown quantitatively by the probability distribution function of current density square $J_z^2$ in Fig.~\ref{fig:jz_pdf}, which we will describe in Section~\ref{energy-decay}.

\subsection{Magnetic topology}
\label{sec:topology}
In this subsection we introduce a method to study the topology of the magnetic field; specifically, we aim to quantify the number of flux tubes on the perpendicular planes so as to make contact with our predictions of Section~\ref{sec:theory}.

Any flux tube crossing a given $xy$ plane will leave a local maximum or minimum point of the flux function on the plane (called the O-point). That is, the cross section of the flux tube on the plane has a structure of a magnetic island. Similarly, the current layers between two merging flux tubes will leave a saddle point on the flux function of the plane (called the X-point).
Both the O-points and the X-points are the critical (or stationary) points of a scalar field --- in this case, the flux function $\psi(x,y)$ on the plane. 
In principle, on each $xy$ plane, the number of O-points should be the same as that of X-points. 
They help characterize in a quantitative way the topology of the magnetic field and can be identified using the technique which we describe next.

We identify the X/O points using the method of Hessian matrix, following previous work~\citep{servidio2009magnetic,servidio2010}. On a given $xy$ plane, we first identify all the critical points of the magnetic potential, defined by $\mathbf{\nabla_\perp}\psi=0$, where the perpendicular magnetic field $\mathbf{B}_\perp=0$. Then we numerically calculate the second-order partial derivatives of the flux function $\psi$ at the critical points, constructing the Hessian matrix
\begin{equation}
    H =
\begin{pmatrix} 
\partial_{xx} \psi & \partial_{xy} \psi \\
\partial_{yx} \psi & \partial_{yy} \psi 
\end{pmatrix}
\end{equation}
and computing its determinant and eigenvalues. 
If $H$ is positive definite at point $(x,y)$, then the point is an O-point with local minimum; and {\it vice-versa}: 
a point with negative definite $H$ is an O-point with local maximum. If $H$ has one positive and one negative eigenvalues, such point is a saddle point, called X-point. 
Rarely, the determinant of $H$ is zero; the locations where this happens are called degenerate critical points.

\begin{figure}
\centering
\includegraphics[width=2.5 in]{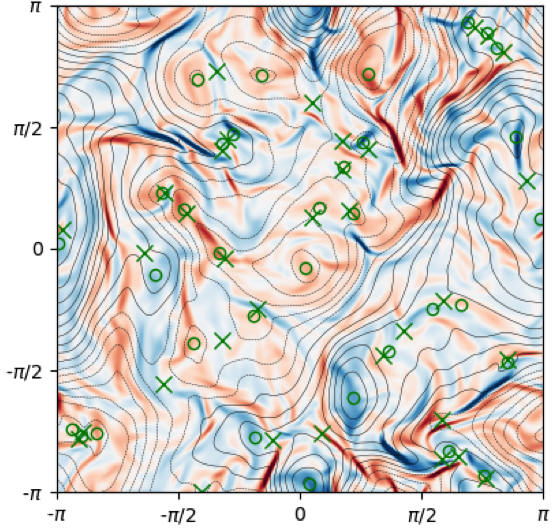}
\caption{Current density $J_z$ (colors) and magnetic flux $\psi$ (contours) on the $xy$ plane at $z=L_z$, at $t=100 \tau_A$ for the run with $S_0=1250$. Green circles (`{\Large $\circ$}' symbols) identify the positions of O-points; Green crosses (`$\times$' symbols) mark the positions of X-points.} 
\label{fig:xo_vis}
\end{figure}

We apply this diagnostic to time slices throughout the simulation, thus obtaining the time evolution of the number of X/O-points, $N_{\perp}(t)$. 
For each time slice, we examine  eight $xy$ planes spread evenly along the $z$-direction. 
An example is shown in Fig.~\ref{fig:xo_vis}. 
Lastly, the number of X/O-points is averaged over the eight planes (all $xy$ planes should be statistically equivalent).
\begin{figure}
    \centering
    \includegraphics[width=3 in]{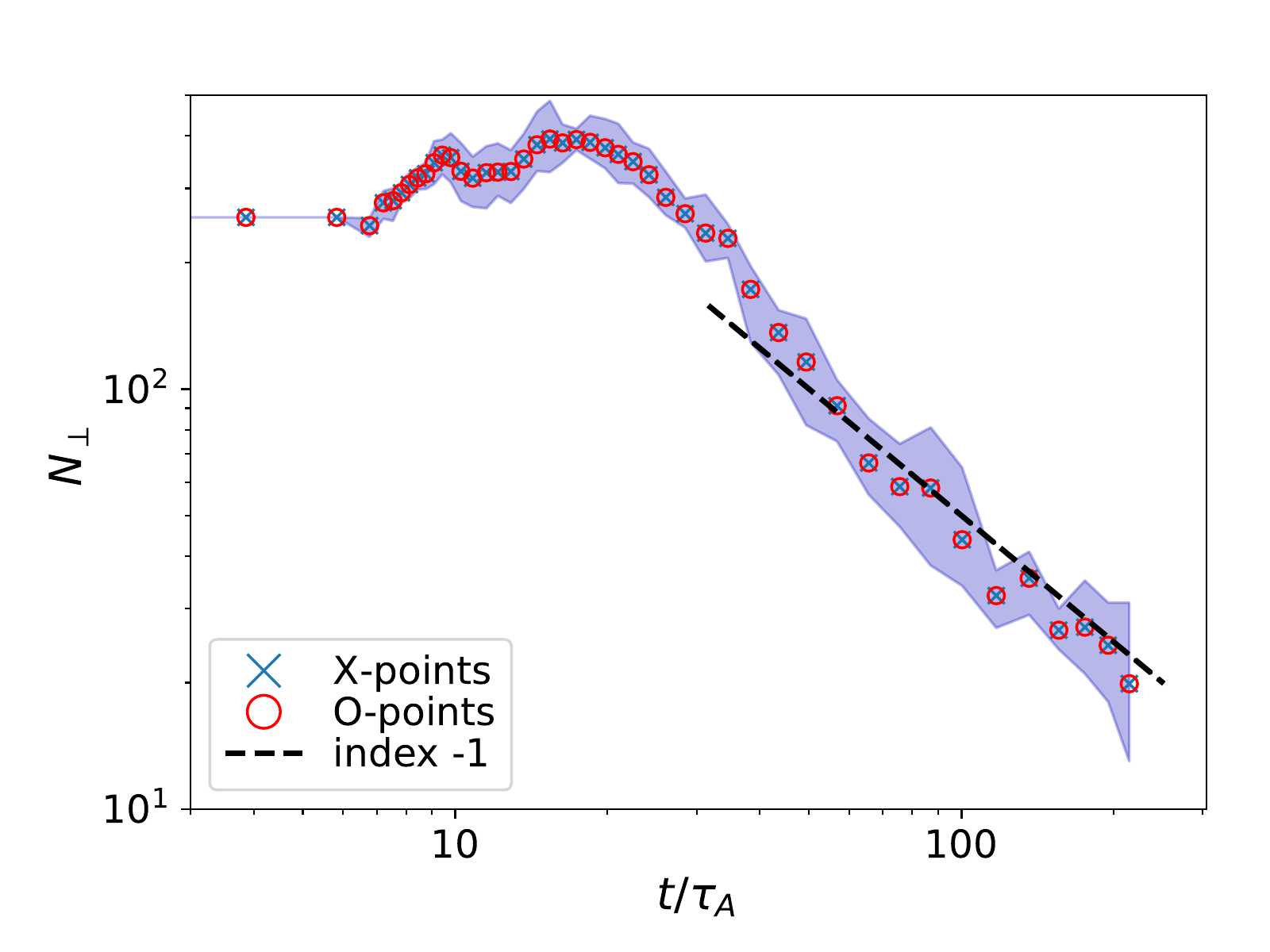}
    \caption{Time evolution of the number of X and O points, $N_{\perp}$, for $S_0=1250$. The data points show the values of the number of X/O-points averaged over eight $xy$ planes for each timeslice. The shadowed area shows the variation between different $xy$ planes. A $t^{-1}$ power law, predicted in Section~\ref{sec:theory}, is shown for reference.}
    \label{fig:Nxy_t}
\end{figure}

Fig.~\ref{fig:Nxy_t} shows the time evolution of $N_{\perp}$ for the run with $S_0=1250$. The data points are the number of X/O-points averaged over eight $xy$ planes and the shadowed area shows the variation between different $xy$ planes.
Initially there are $16\times16=256$ X/O points, as specified by our initial condition. 
During the transient stage (between $\sim 6\tau_A - 30\tau_A$), the number of X/O points first increases by roughly a factor of 2, and then rapidly decreases.
During this stage, the increase in the number of X/O-points is correlated to the deformation of flux tubes and the formation of current sheets (Fig.~\ref{fig:2d_vis} and Fig.~\ref{fig:curret_layers}). This indicates the formation of reconnection sites at multiple places and leads to rapid magnetic energy dissipation, as we will discuss in Section~\ref{energy-decay}.

After the transient stage, the decay of $N_{\perp}$ slows down and scales with time as $N_{\perp} \sim t^{-1}$, as it predicted by our hierarchical model [Eq.~\eqref{eq:number_scalings}].
This suggests that the late-time evolution of the system is indeed governed by the merger of magnetic flux tubes.

\subsection{Energy decay}
\label{energy-decay}
From the visualizations and the number of magnetic structures, we can see that the system first goes through a transient stage of developing turbulence. This is followed by a prolonged stage of decaying turbulence, with the formation of progressively larger magnetic structures.
The behavior of magnetic energy evolution, as we will now describe, is consistent with this physical picture.

\begin{figure}
    \centering
    \includegraphics[width=0.6\textwidth]{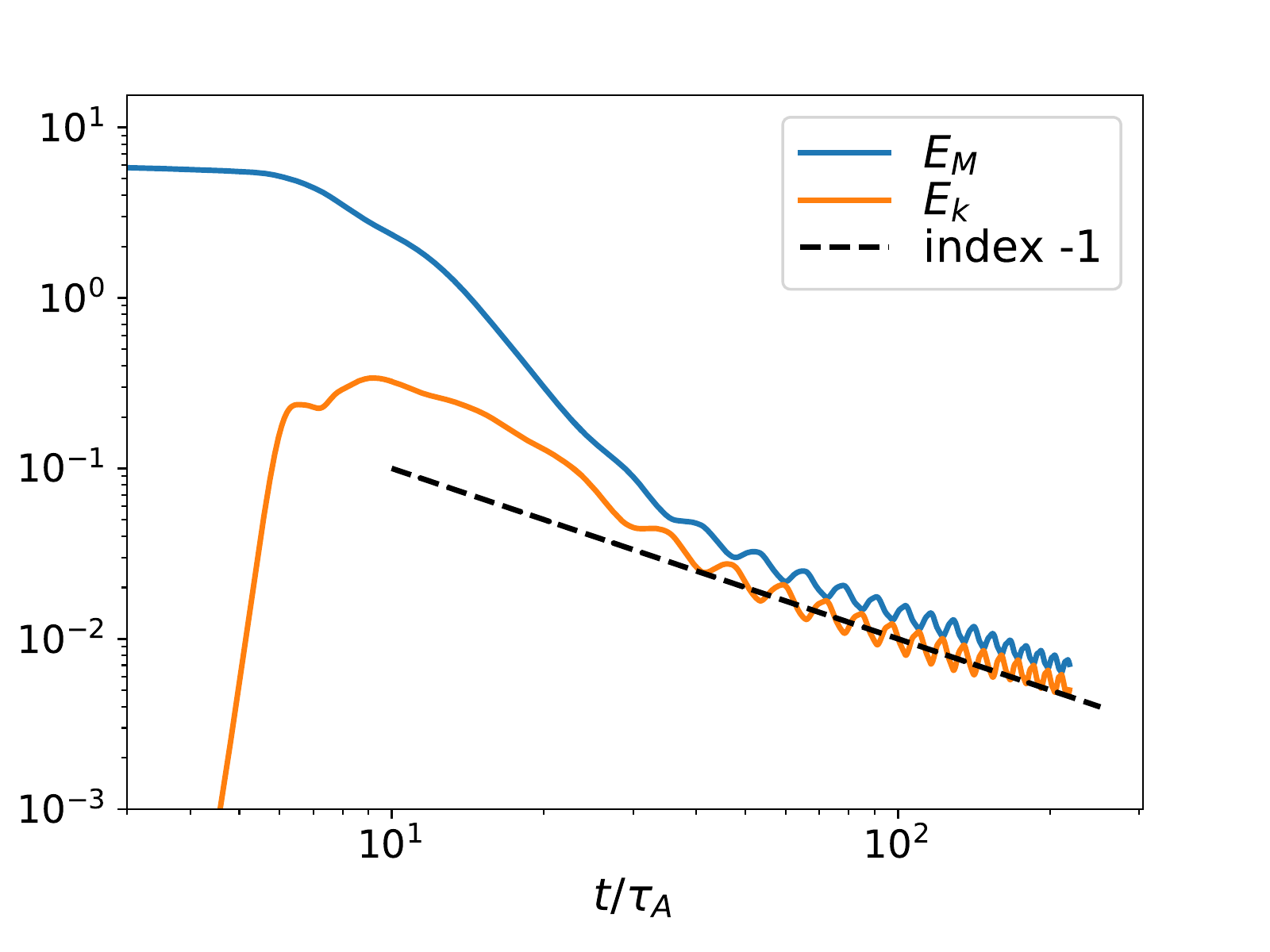}
    \caption{Time evolution of total magnetic energy ($E_M$) and kinetic energy ($E_K$) for $S_0=1250$. A $t^{-1}$ power law is shown for reference, in accordance with our prediction, Eq.~\eqref{eq:field_scalings}.}
    \label{fig:energy_decay}
\end{figure}

Fig.~\ref{fig:energy_decay} shows the time evolution of magnetic and kinetic energies for the run with $S_0=1250$.
Rapid magnetic energy decay occurs during the transient stage (around 6$\tau_A$ to $30\tau_A$), corresponding to the increase, followed by rapid decrease, of the number of X/O points (Fig.~\ref{fig:Nxy_t}). Also, it corresponds to the stage, as we observe from the visualizations, of increasing complexity with the breaking of flux tubes and the emergence of finer structures.
Therefore, we attribute the rapid magnetic energy dissipation during this stage to the formation of a large amount of reconnection sites with localized current sheets among the twisted and deformed structures in this complex system.  
After this stage ($t\gtrsim 30\tau_A$), the magnetic energy decay clearly follows a $t^{-1}$ scaling, as predicted by our hierarchical model, Eq.~(\ref{eq:field_scalings}). 
It coincides with the stage when the number of X/O points exhibits the same scaling, $N_{\perp} \sim t^{-1}$ (Fig.~\ref{fig:Nxy_t}), also in accordance with our predictions, Eq.~(\ref{eq:number_scalings}). 
These numerical confirmations of our predictions suggest that indeed the key process underlying the evolution of our system at this stage is the merger of flux tubes, which appears to proceed as we envisionsed in Section~\ref{sec:theory}.

\begin{figure}
    \centering
    \includegraphics[width=0.6\textwidth]{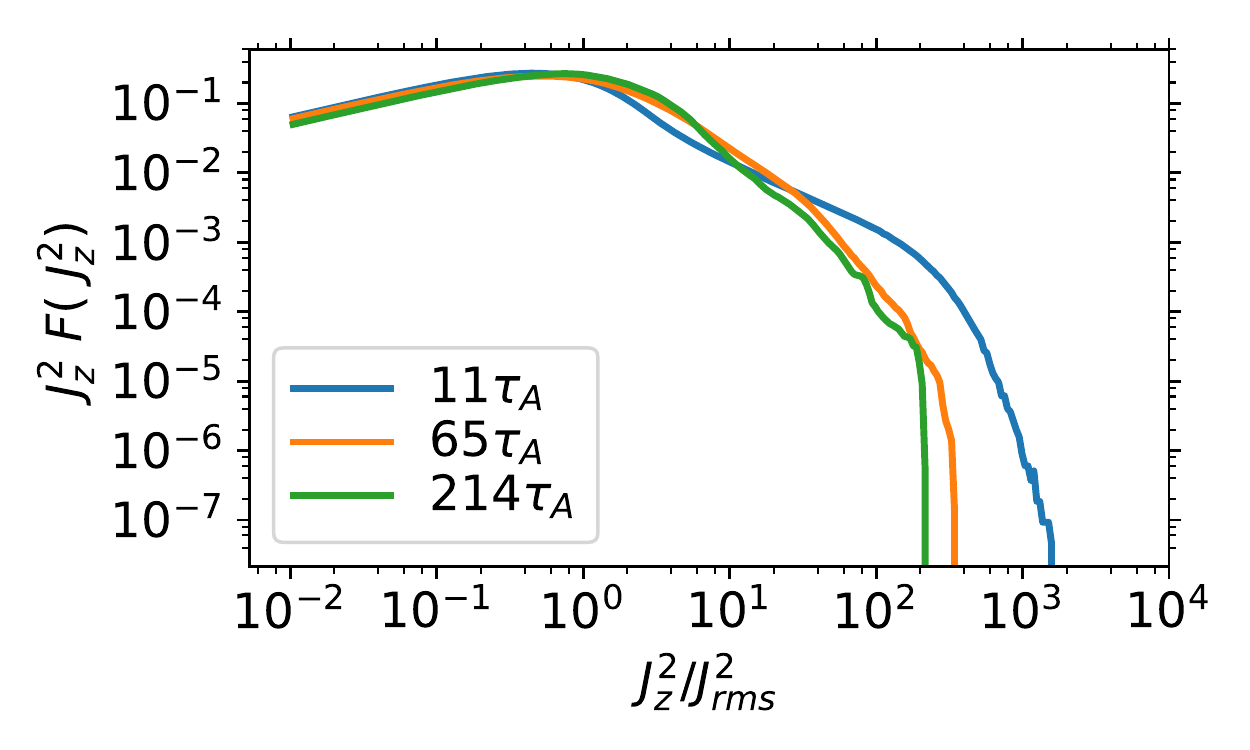}
    \caption{Compensated probability distribution function (PDF) of $J_z^2/J_{\rm rms}^2$ for $S_0=1250$ at three moments of time. The tail at large $J_z^2/J_{\rm rms}^2$ for the PDF at $t=11\tau_A$ indicates a greater number of localized current sheets in the transient stage.}
    \label{fig:jz_pdf}
\end{figure}

Further evidence to support the claims on enhanced energy dissipation during the transient stage is provided in Fig.~\ref{fig:jz_pdf}, which shows the compensated probability distribution function (PDF) of current density square, $J_z^2$, normalized to its mean square, $J_{\rm rms}^2$, at different moments of time 
[in resistive RMHD with constant resistivity, the magnetic energy dissipation (ohmic heating) rate per unit volume is proportional to $J_z^2$].
We arbitrarily choose one snapshot at the transient stage ($t=11 \tau_A$) and two from the self-similar stage ($t=65 \tau_A$ and $t=214 \tau_A$).
The PDF of the two later snapshots (during the self-similar stage) are similar, while the PDF at $t=11 \tau_A$ has a clear tail at the high current density end. 
This suggests that, during the transient stage, there are more localized current sheets facilitating the rapid dissipation of magnetic energy.

The kinetic energy, starting from zero, rapidly increases as the flux tubes start to interact. 
It then traces the magnetic energy and undergoes the transient rapid-decay stage and $t^{-1}$ self-similar stage. Comparing the vorticity density $\omega_z$ and the current density $J_z$ on the same $xy$ planes at the same moments of time (Fig.~\ref{fig:2d_vis}), we find that the flow in the system is dominated by the reconnection outflows with the Alfv\'en speed based on the perpendicular magnetic field. Therefore, when the magnetic energy decays through the merger of flux tubes, the kinetic energy follows magnetic energy and decays as $t^{-1}$.
During the self-similar stage, a small oscillating exchange between magnetic energy and kinetic energy is observed, which we will discuss in Section~\ref{sec:spectra}.

The rapid magnetic energy decay that we observe in the transient stage bears similarities to that observed in the numerical study of magnetized jets from active galaxies~\citep{Neill2012,alves2018}. A jet can be viewed as a screw-pinch/single flux tube that is unstable to current-driven instabilities, such as the kink instability.
Such instabilities distort the flux tube and create localized transient current sheets, enhancing the dissipation of magnetic energy. 
In our system with a strong guide field, we do not expect kink-type instabilities (see discussion in Appendix~\ref{sec:kink}). 
However, the configuration of a large ensemble of flux tubes has more degrees of freedom to form reconnection sites via their twisting and deformation.

\subsection{Spectra and related quantities}
\label{sec:spectra}
\begin{figure}
    \centering
    \includegraphics[width=0.6\textwidth]{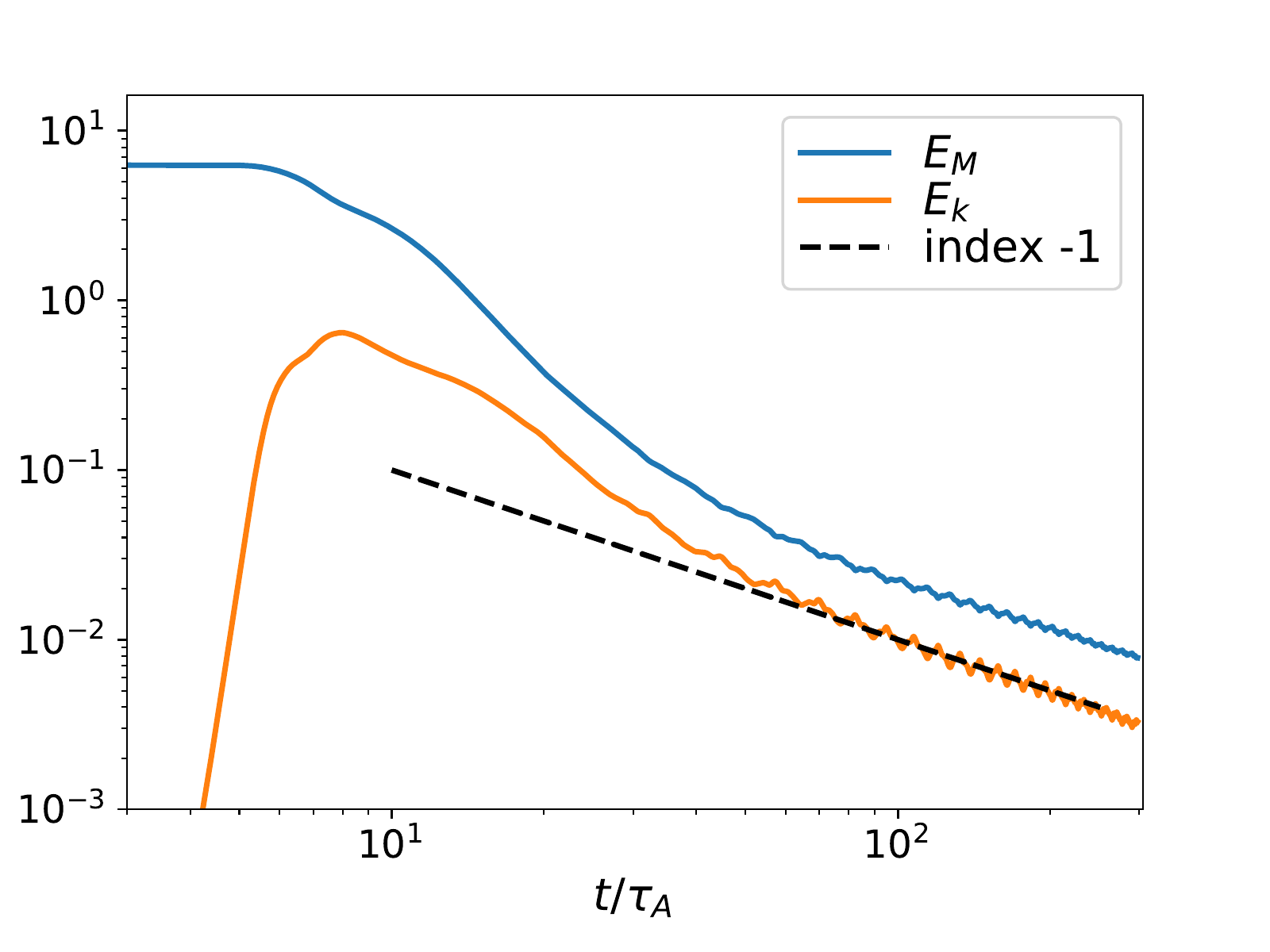}
    \caption{Time evolution of total magnetic ($E_M$) and kinetic ($E_K$) energies for the run with hyper-dissipation.}
    \label{fig:energy_decay_hyper}
\end{figure}

We now proceed to investigate the time evolution of the magnetic and kinetic energy spectra. In order to obtain the clearest spectra possible, we analyze the run with hyper-dissipation instead of normal (Laplacian) diffusivity.
For reference, the time evolution of energies for this run is shown in Fig.~\ref{fig:energy_decay_hyper}. Comparison with Fig.~\ref{fig:energy_decay} shows qualitatively similar behavior between the two runs, although two differences stand out: (i) in the hyper-dissipative run the kinetic energy remains about a factor of 2 below the magnetic energy, whereas in the $S_0=1250$ resistive run they have comparable magnitudes in the self-similar part of the evolution (i.e., for $t/\tau_A \gtrsim 40$); and (ii) the amplitude of the oscillations in the late evolution of the energies is significantly reduced in the hyper-dissipative case. 
We will first discuss the energy spectra and then turn to the reasons for these differences at the end of this subsection.

Generally speaking, the magnetic energy in the system is contained not only in magnetic flux tubes but also in Alfv\'en wave packets launched by the convective motion of flux tubes in the perpendicular plane at the scale of flux tubes.
Magnetic energy contained in flux tubes is transferred to larger scales through the flux-tube mergers, while that contained in Alfv\'en wave packets cascades to smaller scales through nonlinear interactions.  
In the Alfv\'en wave packets, the kinetic energy is in equipartition with the magnetic energy, cascading to smaller scales. A $k^{-3/2}$ spectrum is expected for both the magnetic and the kinetic energy as a result of the turbulent direct cascade~\citep{bolyrev2006}.
There is another ingredient in the system that can potentially contribute significantly to the magnetic and kinetic energy spectrum --- the thin current sheets between merging flux tubes.
The sharp magnetic field reversals at the current sheets are expected to yield a $k^{-2}$ magnetic energy spectrum~\citep{burgers1948mathematical}. 
The Alfv\'enic outflows from reconnection layers, whose spatial profile is derived in \citet{loureiro2013plasmoid}, are expected to yield a flat kinetic energy spectrum.
That is, the thin current sheets in the system can potentially steepen the turbulent magnetic energy spectrum, and flatten the turbulent kinetic energy spectrum.

\begin{figure}
    \centering
    \includegraphics[clip,width=1.0\textwidth]{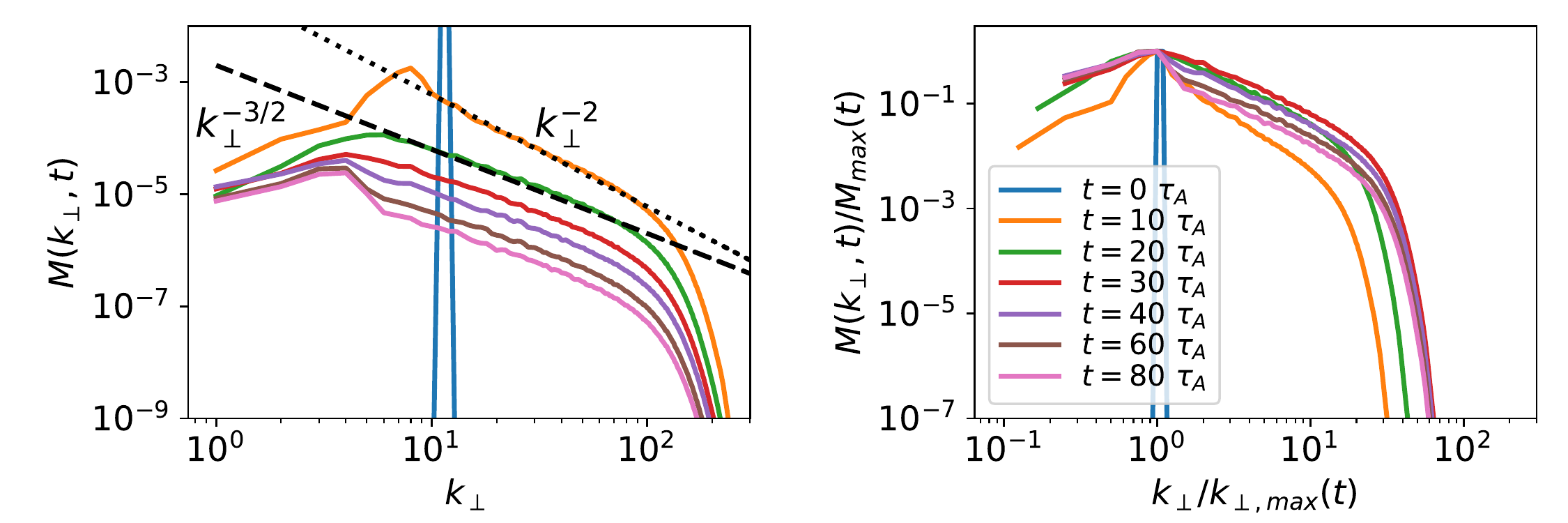}
    \caption{Raw (left) and normalized (right) perpendicular magnetic power spectra, for simulation with hyper-dissipation. Dotted and dashed lines indicate reference slopes of $k_\perp^{-2}$ and $k_\perp^{-3/2}$, respectively.}
    \label{fig:Spectra}
\end{figure}

\begin{figure}
    \centering
    \includegraphics[clip,width=0.6\textwidth]{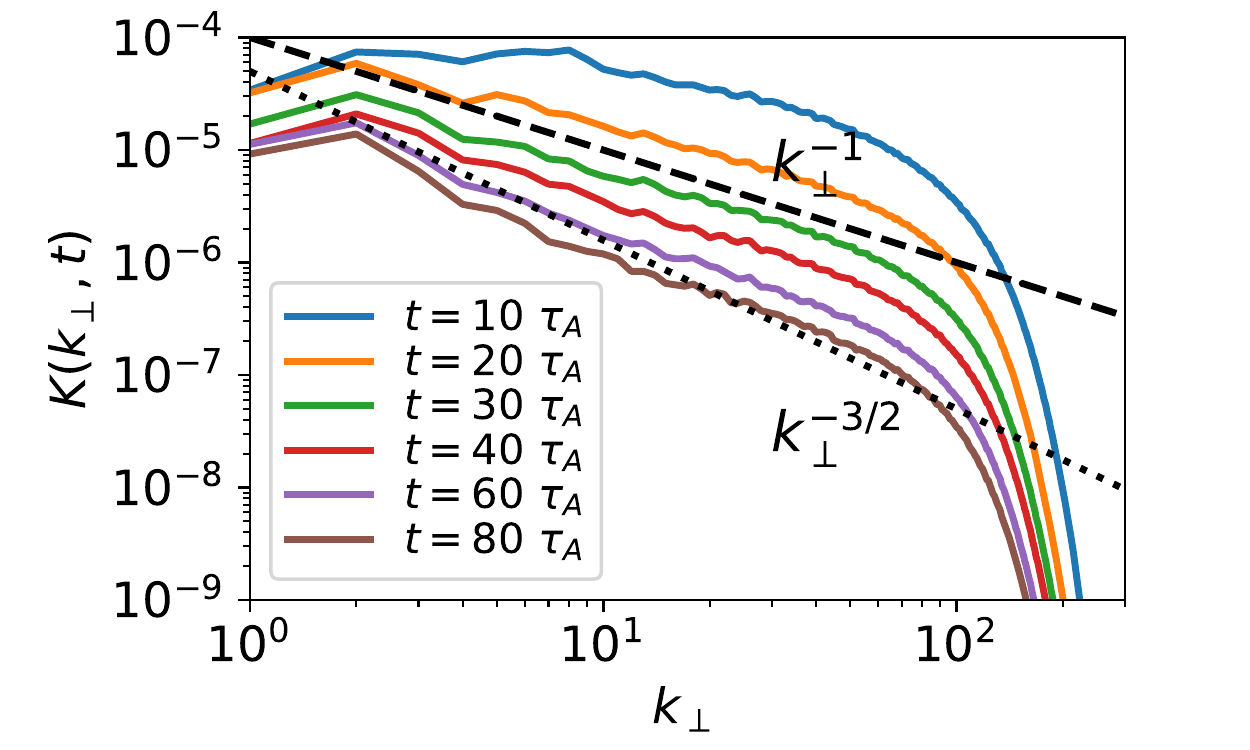}
    \caption{Perpendicular kinetic power spectra for the simulation with hyper-dissipation. $k_\perp^{-1}$ and $k_\perp^{-3/2}$ slopes are shown for reference.}
    \label{fig:Spectra_kin}
\end{figure}

With these considerations borne in mind, let us analyze the numerical results.
The perpendicular magnetic energy spectrum $M(k_\perp,t)$ at different moments of time is shown in Fig.~\ref{fig:Spectra} (left panel). The initial spectrum peaks at the scale corresponding to the cross section of the initial flux tubes. 
As the flux tubes merge, the peak of the spectrum shifts to larger scales, until it becomes limited by the box size. An inertial range appears to the right of the peak, with a power-law exponent $\gamma \approx 2$ at earlier times, and $\gamma \approx 3/2$ at later times. 
The kinetic spectrum, shown in Fig.~\ref{fig:Spectra_kin}, does not have a steady power law.  
At later times, while the magnetic spectrum peaks at $k_\perp=4$, corresponding to the size of islands, the kinetic spectrum peaks at $k_\perp=2$, corresponding to the large-scale motion of the islands. Such motion injects kinetic energy into the system, potentially yielding a power-law inertial range at smaller scales, which we tentatively fit with~$k_\perp^{-3/2}$.
However, at earlier times, the kinetic spectrum follows a shallower power-law $\sim k_\perp^{-1}$. 
While we do not have an explanation for the $-1$ index, we think it is consistent with the observation that at earlier times the system has many localized thin current sheets, where the kinetic energy is dominated by the (Alfv\'enic) outflows.
 The outflows have a spatial profile yielding a flat spectrum as we described above, and act as a source of kinetic energy at all scales.
As the flux tubes merge and grow to larger scales, the thickness of the current sheets also grows and the magnetic reversal becomes less sharp, resulting in a weaker flattening effect on the spectrum.
The kinetic spectrum thus becomes steeper at later times.
At early times (before the turbulence is fully developed), the magnetic spectrum follows a $k_\perp^{-2}$ spectrum resulting from the sharp field reversals~\citep{burgers1948mathematical}. 
At later times, as the turbulent cascade becomes stronger, the $k_\perp^{-2}$ spectrum is replaced by a shallower $k_\perp^{-3/2}$ turbulent spectrum. 
Interestingly, in our 2D study, a $k_\perp^{-2}$ magnetic energy spectrum was maintained throughout the evolution [Fig.~4 in \citetalias{zhou2019magnetic}].
The reason for this difference is that, in the 2D system, the magnetic energy significantly dominates the kinetic energy, making the system less turbulent than its 3D counterpart. 
The direct turbulent cascade is too weak to form a $k^{-3/2}$ spectrum, and therefore, the total magnetic energy spectrum in our 2D system is mostly determined by the current sheets, yielding a $k_\perp^{-2}$ spectrum.

Our analytical model of Section~\ref{sec:theory} leads to the prediction that the long-time evolution of the magnetic energy spectrum should be self-similar, Eq.~(\ref{eq:U_k_t}).
To test this prediction, we show in the right panel of Fig.~\ref{fig:Spectra} the spectra normalized to their instantaneous maximum value, $M_{\rm max}(t)$, as a function of wavenumber normalized to the values $k_{\rm \perp,max}(t)$ at which $M_{\rm max}(t)$ are achieved. We observe that, except for the transient-stage spectrum at $t=10 \tau_A$, the spectra thus normalized approximately collapse onto the same distribution, and that this collapse becomes progressively better as time increases. 
Mathematically, this behavior is expressed as $M(k_\perp,t)=M_{\rm max}(t)\Bar{M}(k_\perp/k_{\rm \perp, max})$, where $\Bar{M}(k_\perp/k_{\rm \perp, max})$ is the static universal distribution to which all the curves collapse after normalization.

\begin{figure}
    \centering
    \includegraphics[width=0.6\textwidth]{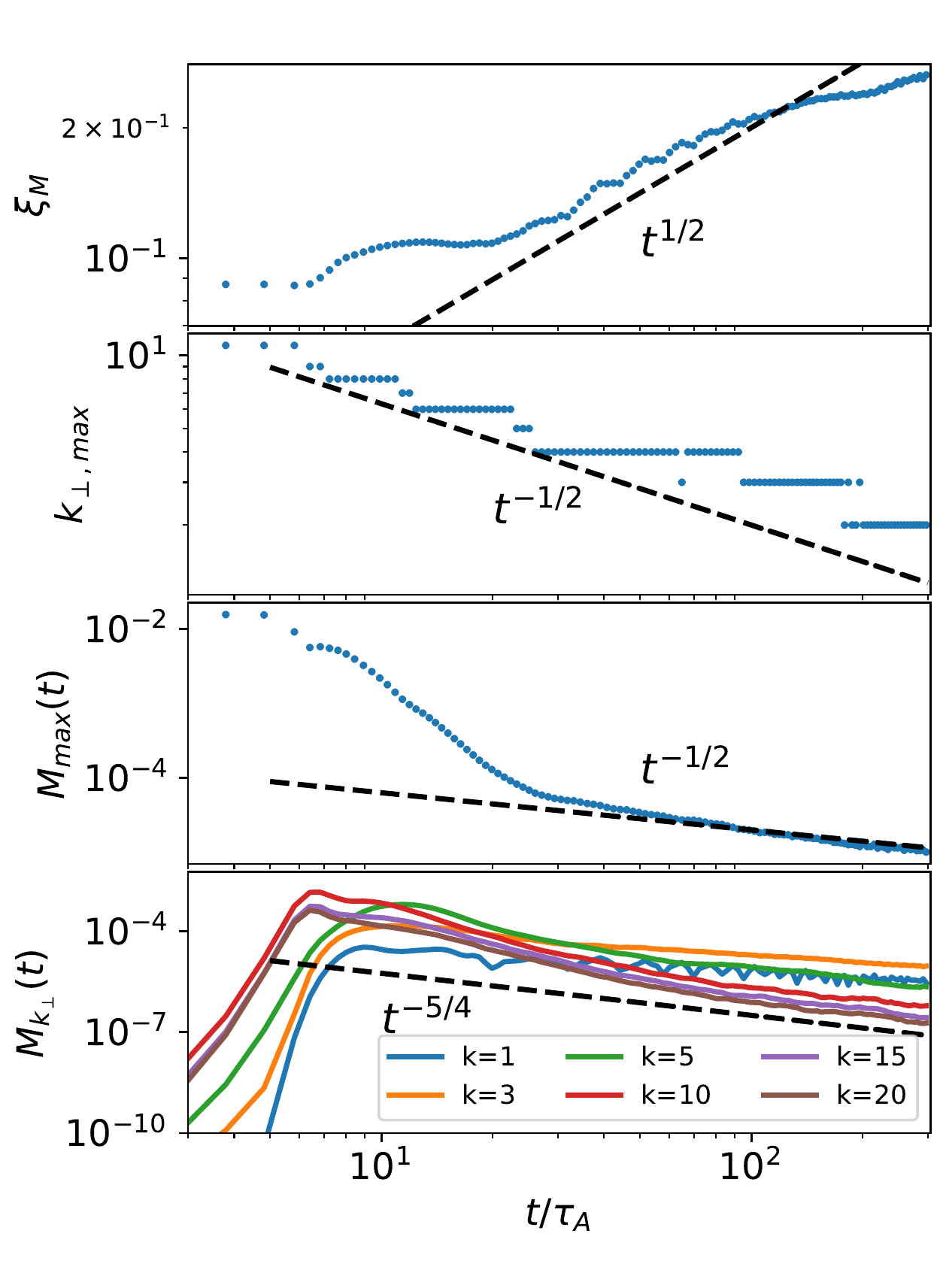}
    \vspace*{-4mm}
    \caption{Time evolution of $\xi_M$ (top), $k_{\rm \perp, max}$ (second), $M_{\rm max}$ (third) and $M_{k_\perp}$ (for selected values of $k$; bottom), for the simulation with hyper-dissipation.}
    \label{fig:kmaxUmax}
\end{figure}

We also observe that after the transient stage, the time evolution of $k_{\rm \perp,max}$ and $M_{\rm max}$ approximately follow the power-law-in-time behavior: $k_{\rm \perp,max}\propto t^{-1/2}$, and $M_{\rm max} \propto t^{-1/2}$ --- see the second and third panels in Fig.~\ref{fig:kmaxUmax}. 
Thus, the magnetic energy spectrum can be expressed as $M(k_\perp, t)\propto t^{-1/2}\Bar{M}(k_\perp t^{1/2})$, in agreement with the prediction of Eq.~\eqref{eq:U_k_t}. 
Furthermore, since the scaling function is itself a power law in the inertial range, $\Bar{M}(k_\perp t^{1/2}) \propto k_\perp^{-3/2}$ ($\gamma=3/2$), according to our prediction, the time evolution of spectral magnetic energy density  at any fixed wavenumber should behave as $M_{k_\perp}(t)\propto t^{-\alpha}$, where $\alpha=(\gamma+1)/2=5/4$. 
This is indeed observed in the bottom panel of Fig.~\ref{fig:kmaxUmax}, where after the transient stage, $M_{k_\perp}(t)$ (for moderate wavenumbers) approaches the $t^{-5/4}$ power law.
At the largest scale the energy keeps building up, explicitly showing the inverse transfer of magnetic energy.   
From the measured scaling laws described above, we conclude that the self-similar solution of the magnetic power spectrum, based on our hierarchical analytical model, is confirmed by our numerical results.

Complementary to the time evolution of $k_{\rm \perp, max}$, the growth of the perpendicular coherence length of our system is illustrated by computing the magnetic energy containing scale, defined as 
\begin{equation}
\label{eq:coherence_length}
    \xi_M(t) \equiv \frac{\int 2\pi k_\perp^{-1}M(k_\perp,t)dk_\perp}{\int M(k_\perp,t)dk_\perp}.
\end{equation}
As shown in the top panel of Fig.~\ref{fig:kmaxUmax}, after the transient stage, $\xi_M$ increases with time roughly following the $t^{1/2}$ scaling, consistent with our model, until the later stage ($t \gtrsim 100 \tau_A$) during which the parallel length of flux tubes becomes limited by the length of the box. 

\begin{figure}
    \centering
    \includegraphics[width=0.6\textwidth]{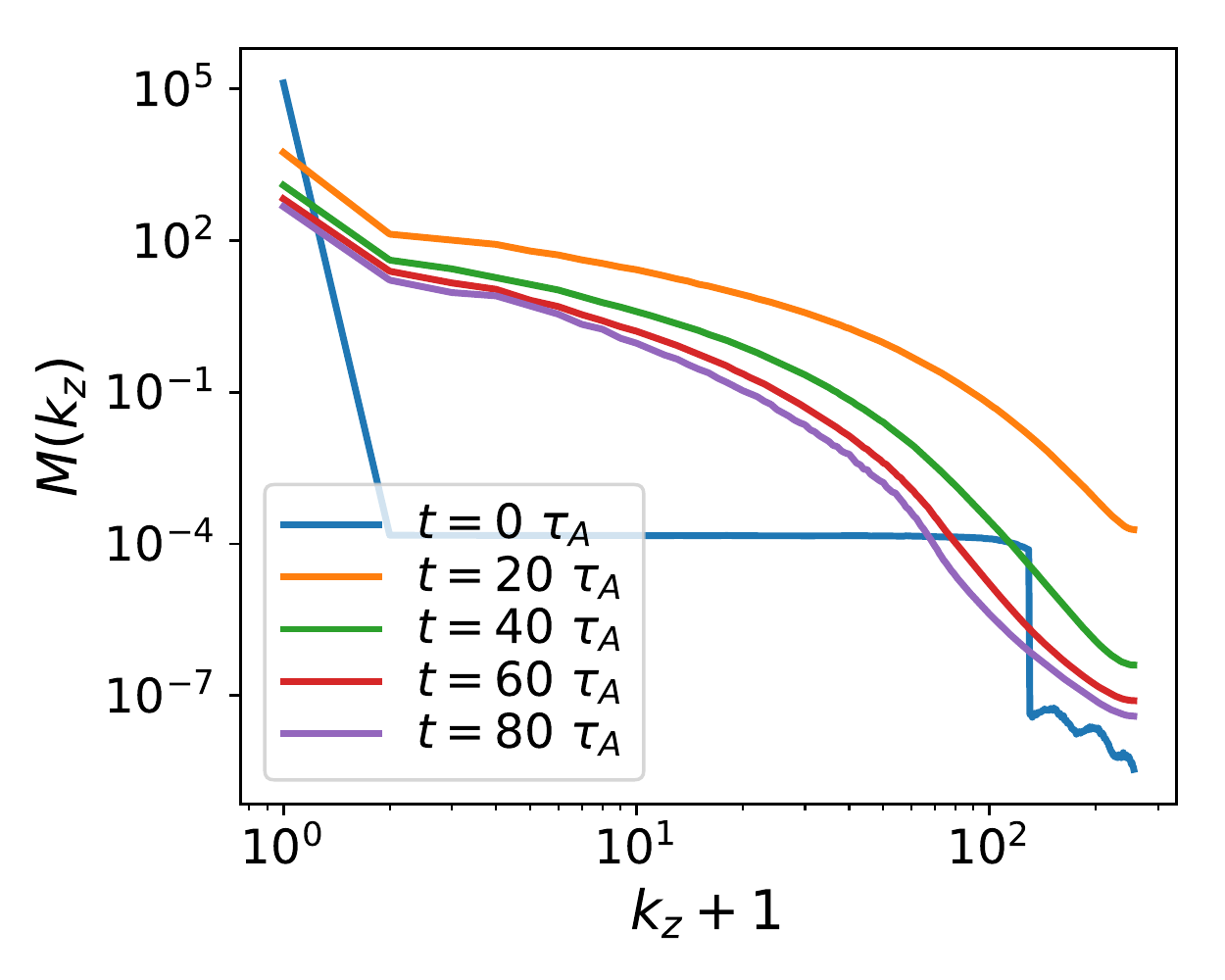}
    \caption{Parallel (to the guide field $\mathbf{B}_g$) magnetic power spectra at different moments of time for the simulation with hyper-dissipation.}
    \label{fig:spec_z}
\end{figure}

To characterize the system's dynamics along the guide field, we show in Fig.~\ref{fig:spec_z} the magnetic power spectrum as a function of~$k_z$. 
A broad energy distribution develops once the flux tubes break in the $z$ direction. 
As flux tubes merge, magnetic energy at high $k_z$ rapidly decreases, indicating the increasing coherence length in $z$ direction.
This is consistent with the measurement of magnetic structure functions (Fig.~\ref{fig:sfs_contour}), showing the rapid growth of the characteristic flux tube, as we explain in Section~\ref{sec:structurefunction}.

Finally, let us address the issue of the oscillatory energy exchange between the magnetic field and the flow. 
From the time evolution of magnetic energy density at different wavenumbers~$k_\perp$ (Fig.~\ref{fig:kmaxUmax}, bottom panel), we see that energy oscillation mostly happens at the smallest~$k_\perp$, corresponding to system-size magnetic flux tubes. 
These tubes cannot merge and grow into larger scales due to the system-size limit (in a sense, they have reached a final `equilibrium' consisting of two flux tubes of opposite polarities). 
Instead, they exchange energy with the flow --- possibly both in the form of Alfv\'en waves, and bouncing motions: background flows can push these flux tubes against each other, leading to their (incompressible) deformation. This increases magnetic field line tension and results in a restitution force leading to an oscillation whose amplitude, therefore, depends on the ratio of the kinetic-to-magnetic energy. 
Since the magnetic energy is larger in the hyper-dissipative case than in the Laplacian resistivity case (simply because hyper-resistivity leads to better conservation of magnetic energy), the hyper-dissipation case has smaller amplitude oscillations.


\subsection{Magnetic structure functions}
\label{sec:structurefunction}
\begin{figure*}
\includegraphics[width=\textwidth,clip]{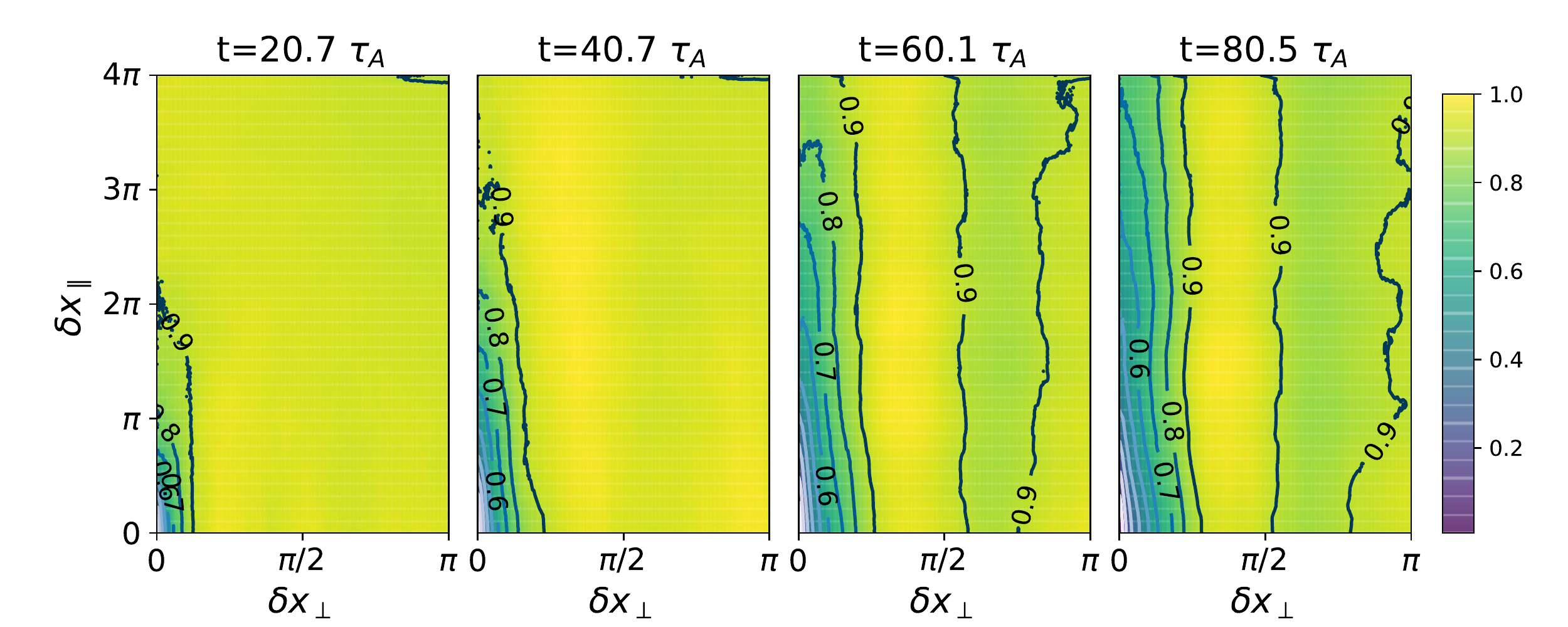}
\vspace*{-5mm}
\caption{Contours of normalized (to its maximum value) magnetic structure function $S_B^{(2)}(\delta x_\perp, \delta x_\parallel)$ at various times for the run with hyper-dissipation. }
\label{fig:sfs_contour}
\end{figure*}

\begin{figure}
    \includegraphics[width=0.45\textwidth]{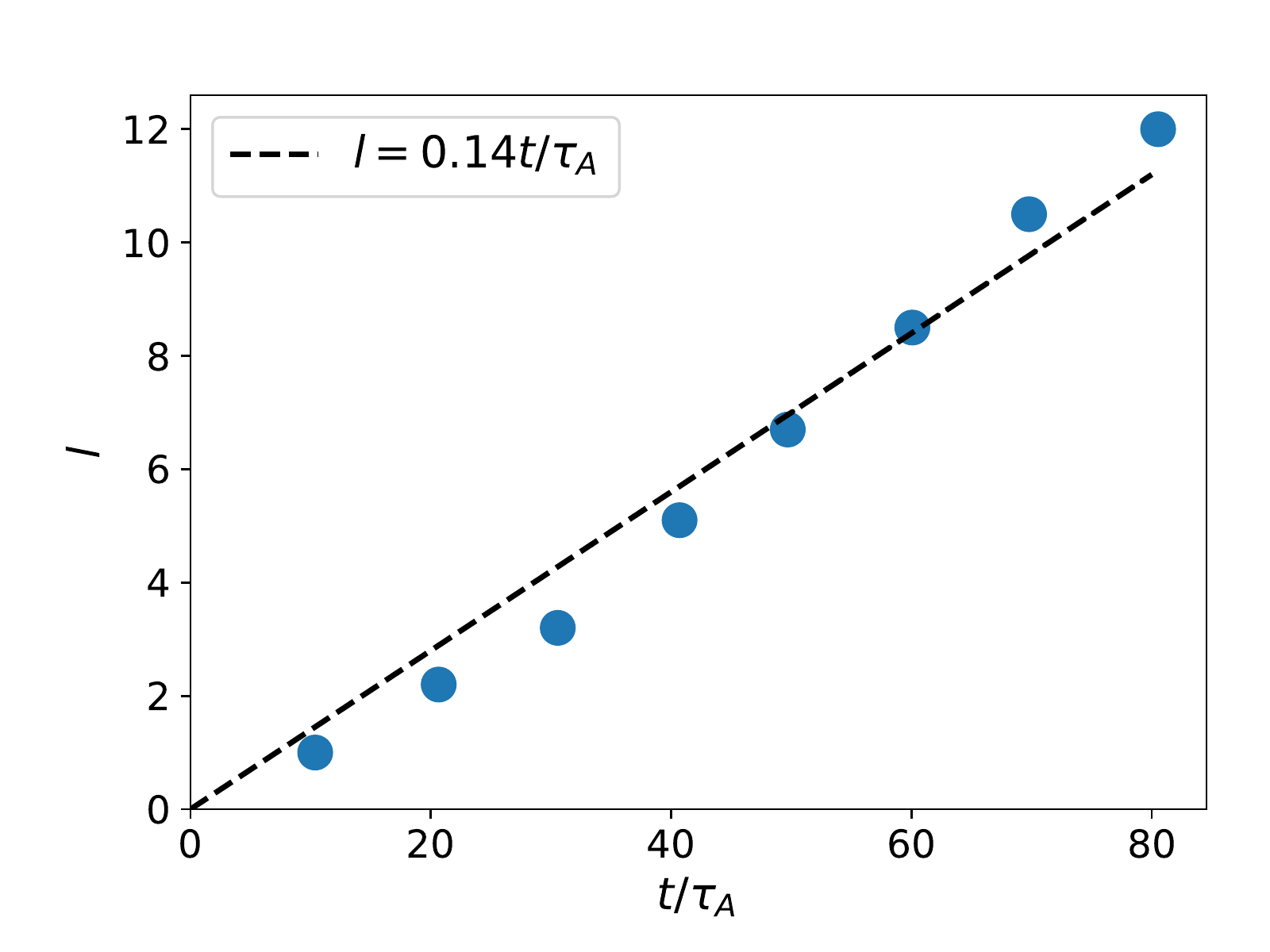}
    \includegraphics[width=0.45\textwidth]{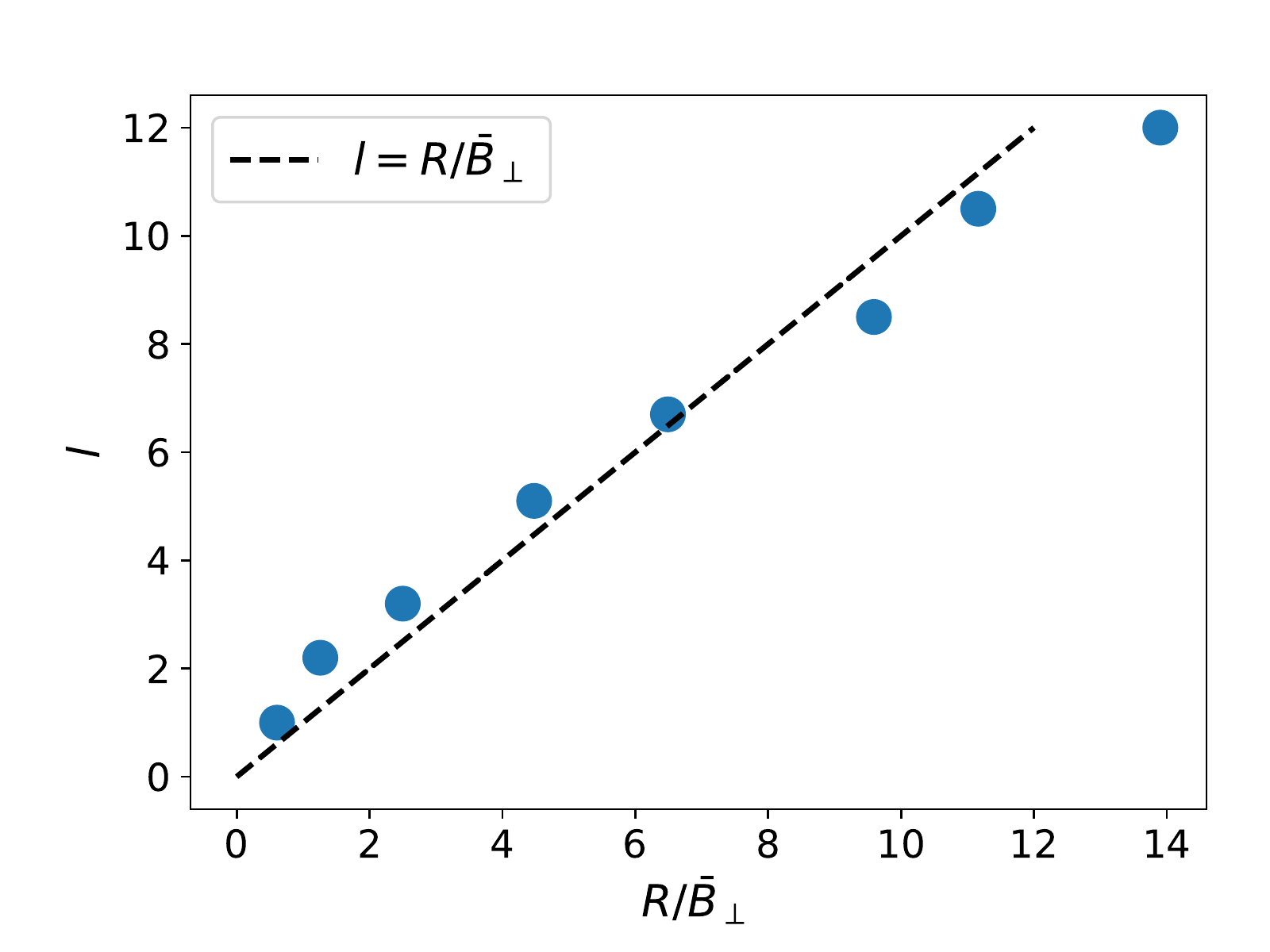}
    \caption{Time evolution of $l$ [left panel, dashed line represents a $t^1$ power law, Eq.~\eqref{eq:parallel_length}] and scaling of $l$ versus $R/\bar{B}_\perp$ [right panel, dashed line represents the critical balance scaing, Eq.~\eqref{eq:critical_balance}] for the run with hyper-dissipation, where $l$ and $R$ are the length and radius of the characteristic flux tube respectively.}
    \label{fig:sfs_cb}
\end{figure}
In order to obtain statistical information on the growth of flux tubes, and to test the assumption of critical balance, we calculate the second-order structure function of the magnetic field $\mathbf{B}(\mathbf{x},t)$, defined as
\begin{equation}
    S_B^{(2)}(\delta \mathbf{x},t)\equiv \braket{|\mathbf{B}(\mathbf{x}_2,t)-\mathbf{B}(\mathbf{x}_1,t)|^2}_{x_1},
\end{equation}
where $\delta \mathbf{x} \equiv \mathbf{x}_2-\mathbf{x}_1$ and the angle brackets indicate the average over the position $\mathbf{x_1}$ in the simulation domain.
We define the local mean field as $\mathbf{B}_m \equiv [\mathbf{B}(\mathbf{x}_1)+\mathbf{B}(\mathbf{x}_2)]/2$ and the local unit magnetic field vector $\hat{\mathbf{b}}_m \equiv \mathbf{B}_m/|\mathbf{B}_m|$ represents its direction. 
These definitions ensure that the structure function is computed with respect to the local mean magnetic field [as it should, if it is to be used to gauge whether the turbulence we observe is critically balanced ~\citep{cho2000,mallet2016}], rather than the global guide field.
The local coordinates ($\delta x_\perp, \delta x_\parallel$) can thus be specified by
\begin{equation}
\begin{aligned}
    &\delta x_\perp \equiv  |\hat{\mathbf{b}}_m \times \delta \mathbf{x}|, \quad \delta x_\parallel \equiv  \hat{\mathbf{b}}_m \cdot \delta \mathbf{x}.
\end{aligned}
\end{equation}

The correlation lengths of the magnetic field parallel and perpendicular to the local mean field, which we interpret as the statistical length and radius of the flux tubes respectively, are thus expressed by the contours of the structure function $S_B^{(2)}(\delta x_\perp, \delta x_\parallel)$ on the ($\delta x_\perp, \delta x_\parallel$) plane.
This is shown in
Fig.~\ref{fig:sfs_contour} (normalized to the instantaneous maximum value of the structure function). 
We then define the characteristic length $l$ and the radius $R$ of the flux tubes using the intersection of the contour line $S_B^{(2)}/S_{\rm B, max}^{(2)}=0.7$ (roughly the largest value at which the contour lines are not affected by noise) with the $\delta x_\parallel$ axis and the $\delta x_\perp$ axis, respectively.
Note that because of periodic boundary conditions, the largest values of $\delta x_\perp$ and $\delta x_\parallel$, and hence maximum radius and length of a flux tube that can be attained in our simulations, are roughly $L_x/\sqrt{2}=\sqrt{2}\pi$ and $L_z/2=4\pi$, respectively.
The growth of the typical flux tube can be clearly observed as the system evolves with time. 
The growth of parallel length is faster than that in the perpendicular direction, which agrees with our discussion in Section~\ref{sec:theory} that the aspect ratio of a flux tube increases with time ($l/R \sim \tilde{t}^{1/2}$).

In order to characterize the time evolution of magnetic flux tubes, we take eight snapshots between $t=10\tau_A$ and $t=80\tau_A$ (roughly every $10\tau_A$). For each snapshot, we compute the values of $l$ and $R$.
The time evolution of $R$ (not shown) roughly follows a $t^{1/2}$ scaling (similar to the time evolution of $\xi_M$ shown in Figure~\ref{fig:kmaxUmax}), whereas the time evolution of $l$ scales linearly with time (left panel of Figure~\ref{fig:sfs_cb}), in accordance with the prediction of Eq.~\eqref{eq:length_scalings}.
We also compute the root-mean-square value of the perpendicular magnetic field, $\bar{B}_\perp$, and test if it satisfies the critical balance scaling, $l/\hat{B}_g \sim R/\bar{B}_\perp$, where 
 $\hat{B}_g=B_g L_\perp/L_\parallel=1$ is the guide field normalized in the same way as used to compute the structure function.
We plot $l$ versus $R/\bar{B}_\perp$ for these eight snapshots in the right panel of Figure~\ref{fig:sfs_cb}.
As seen, the critical balance condition is indeed verified (except for the last snapshot, whereupon the parallel length of flux tubes has reached the length of the simulation box, and the critical balance scaling is thus expected to break). 
This confirms our assumption made in Section~\ref{sec:theory} that in our system of merging flux tubes, the aspect ratio of the flux tubes, $l/R$, is determined by critical balance.

\subsection{Energy transfer function}
\label{sec:transferfunction}
In this subsection we use shell-to-shell energy transfer functions to quantify the transfer between magnetic and kinetic energy at different scales. This enables us to identify direct and inverse energy transfer in our simulations. 
This technique has been used in studies of MHD and kinetic turbulence~[e.g., \cite{alexakis2005shell,told2015multiscale}], and we made modifications for its use in the RMHD system. 

We define the shell-filtered variables as 
\begin{equation}
    \mathbf{B}_\perp^K(\mathbf{x}_\perp,z) = \sum_{K < k_\perp \leq K+1} \mathbf{B}_\perp(\mathbf{k}_\perp,z) e^{i\mathbf{k}_\perp \cdot\mathbf{x}_\perp}
\end{equation}
and
\begin{equation}
    \mathbf{u}_\perp^K(\mathbf{x}_\perp,z) = \sum_{K < k_\perp \leq K+1} \mathbf{u}_\perp(\mathbf{k}_\perp,z)e^{i\mathbf{k}_\perp \cdot\mathbf{x}_\perp}.
\end{equation}
Here $\mathbf{x}_\perp$ is the projection of the position vector onto the $x$-$y$ plane and the Fourier and inverse Fourier transforms are only performed in the $x$- and $y$- directions. The shell-filtered variables contain only the magnetic and velocity fields within a certain range of perpendicular scales. The shell transfer functions are obtained by calculating the evolution equation of energy in each shell averaged over the $z$ direction. 
We are interested in the time evolution of magnetic energy in each shell $K$:
\begin{multline}
    \partial_t  \left\langle \int  d^2 \mathbf{x}_\perp  \frac{1}{2}(\mathbf{B}_\perp^K)^2  \right\rangle_z =  \\ \left\langle \int  d^2 \mathbf{x}_\perp \sum_Q\left[ \mathbf{B}_\perp^K \cdot (\mathbf{B}_\perp \cdot \mathbf{\nabla}_\perp) \mathbf{u}_\perp^Q -  \mathbf{B}_\perp^K \cdot (\mathbf{u}_\perp \cdot \mathbf{\nabla}_\perp) \mathbf{B}_\perp^Q \right] - \eta | \nabla_\perp \mathbf{B}^K|^2  \right\rangle_z.
\end{multline}
The transfer of magnetic energy from shell $Q$ to shell $K$ is
\begin{equation}
    T_{bb}(Q,K) = - \langle \int  d^2 \mathbf{x}_\perp \mathbf{B}_\perp^K \cdot (\mathbf{u}_\perp \cdot \mathbf{\nabla}_\perp) \mathbf{B}_\perp^Q \rangle_z,
\end{equation}
and the transfer of kinetic energy in shell $Q$ to magnetic energy in shell $K$ is
\begin{equation}
    T_{ub}(Q,K) = \langle \int  d^2 \mathbf{x}_\perp \mathbf{B}_\perp^K \cdot (\mathbf{B}_\perp \cdot \mathbf{\nabla}_\perp) \mathbf{u}_\perp^Q \rangle_z.
\end{equation}

\begin{figure}
    \centering
    \includegraphics[width=0.45\textwidth]{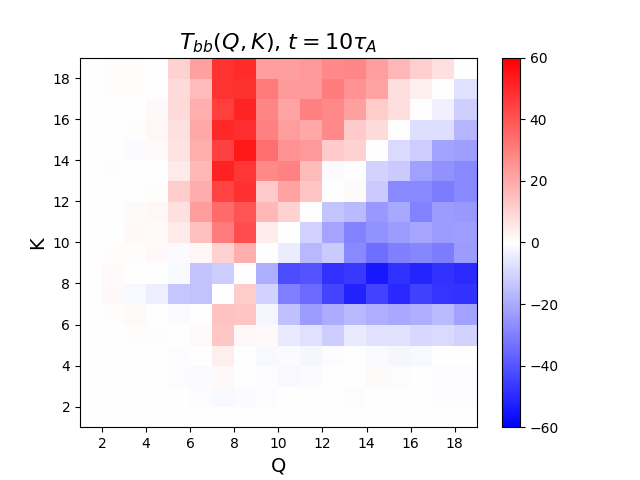}
    \includegraphics[width=0.45\textwidth]{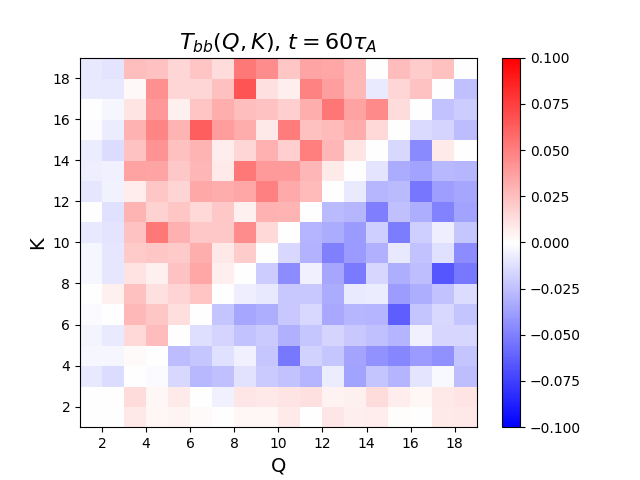}
    \includegraphics[width=0.45\textwidth]{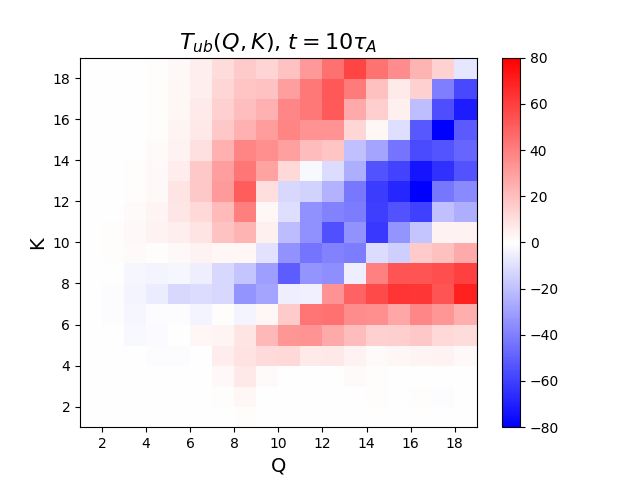}
    \includegraphics[width=0.45\textwidth]{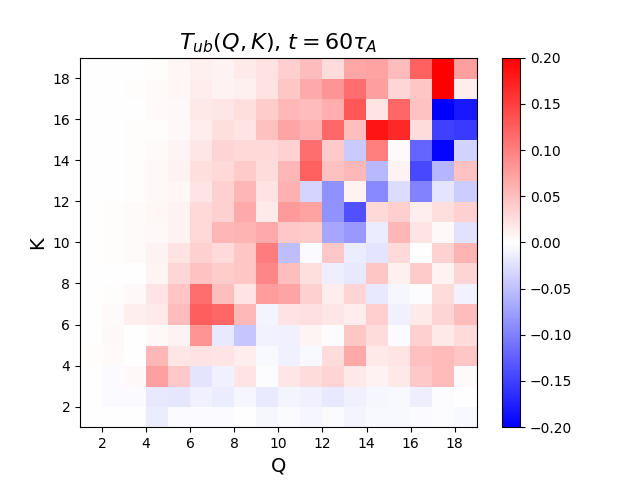}
    \caption{Averaged energy transfer from shell $Q$ to shell $K$ at $t=10\tau_A$ (first stage) and $t=60\tau_A$ (second stage) for the run with hyper-dissipation. Top: magnetic-to-magnetic transfer $T_{bb}(Q,K)$. Bottom: kinetic-to-magnetic transfer $T_{ub}(Q,K)$. }
    \label{fig:trans_func_contour}
\end{figure}

We perform the energy transfer function diagnostic on the results of the run with hyper-dissipation. We arbitrarily choose one snapshot during the first stage of developing turbulence ($t=10\tau_A$) and another during the second stage of decaying turbulence ($t=60\tau_A$) to illustrate the behaviour of shell-to-shell energy transfer, as shown in Figure~\ref{fig:trans_func_contour}.
To guide the reader, we note that red color below (above) the diagonal in these plots denotes an inverse (direct) energy transfer; and vice-versa for the blue color.
We first focus on the contour plot of $T_{bb}(Q,K)$ at $t=10\tau_A$, shown in the top-left panel.
At this moment, the magnetic energy spectrum shown in Figure~\ref{fig:Spectra} peaks at $k_\perp = 8$.
There is a strong direct energy transfer from shells $Q=7,8$ to shells $K>Q$, indicating the non-local direct transfer characterizing the formation of thin current sheets (consistent with the  $k_\perp^{-2}$ magnetic energy spectrum as we explained earlier, shown in the left panel of Figure~\ref{fig:Spectra}).
There is also a weak but statistically significant inverse transfer of magnetic energy from shells $Q=7,8$ to shells $K=5-7$ (corresponding to the peak of magnetic spectrum). This suggests that the inverse transfer happens at the scale of flux-tubes and is caused by flux-tube merger. 
Due to the development of turbulence, $T_{bb}$ is dominated by the direct transfer from small-$Q$ shells to large-$K$ shells overall. 
At this stage, there is also a transfer of magnetic energy to kinetic energy close to the diagonal, as shown in the contour plot of $T_{ub}$ in the bottom-left panel, consistent with the formation of plasma flow.

At $t=60\tau_A$, flux tubes have reached larger scales (with a peak at around $k_\perp=4$). 
Shells $K,Q>3$ are dominated by the strong direct transfer between a wide range of scales.
However, although relatively weak in strength, there is also a finite inverse magnetic energy transfer from shells $Q>3$ to shells $K=1-2$ shown by the red horizontal band along the lower boundary of the contour plot of $T_{bb}$ in the top-right panel.
From the $T_{ub}$ plot in the bottom-right panel, we observe a mild transfer from magnetic energy in shells $K=1,2$ to kinetic energy in shells $Q>4$.
For $K>3$, kinetic energy in shells $Q$ is transferred to magnetic energy in shells $K$ if $Q<K$, while for $Q>K$, the transfer between kinetic and magnetic energy happens in both directions.

In summary, the energy transfer diagnostics paint a complex picture that is consistent with the analytical model and numerical results discussed earlier.
Both inverse and direct energy transfer occurs: the former driven by flux-tube coalescence; the latter by the turbulent dynamics that arises. While the direct energy transfer is dominant, it is the inverse transfer that progressively delivers energy to larger and larger scales until reaching the simulation domain size.

\section{Conclusion}
\label{sec:conclusions}
In this work, we investigate the inverse transfer of magnetic energy in 3D MHD turbulence, and propose the merger of magnetic flux tubes through magnetic reconnection as its underlying physical mechanism.
An analytical model is developed in the 3D RMHD framework, based on the hierarchical merger of flux tubes.
Perpendicular to the strong guide field, the dynamics are found to be identical to those in the analogous 2D system. The perpendicular magnetic field energy decays as $\tilde{t}^{-1}$ and the perpendicular correlation length grows as $\tilde{t}^{1/2}$, where $\tilde{t}$ is the time normalized to the reconnection time scale. 
Parallel to the guide field, the dynamics are determined by critical balance, setting the aspect ratio of the flux tubes. The power-law time evolution of physical quantities leads to the self-similar evolution of the perpendicular magnetic spectrum $M(k_\perp,\tilde{t})\propto\tilde{t}^{-\alpha}k_\perp^{-\gamma}$, where $2\alpha=\gamma+1$.

Our analytical model is confirmed by direct numerical solution of 3D incompressible RMHD equations. The initial equilibrium (without plasma flow) is set up with an ensemble of volume-filling flux tubes of alternating polarities aligned with the guide field. Given a small perturbation, the system undergoes a two-stage evolution. 
The system first enters a transient stage of developing turbulence, during which the flux tubes break in the parallel direction, leading to the development of turbulence and enhanced magnetic energy dissipation. A prolonged self-similar stage of decaying turbulence ensues, during which the evolution of various quantities is captured by our analytical model and the assumption of critical balance is confirmed by the measurement of 2nd-order magnetic structure function.

Our 3D system shares similarities and differences with its 2D counterpart. 
The presence of a strong guide field (underlying the RMHD formalism that we employ in our simulations) implies that the merger of flux tubes in the perpendicular planes is identical in 3D and in 2D. 
That is, in both cases the mergers are well described by rules stemming from the conservation of mass and poloidal flux (or magnetic potential). 
A novel aspect of our work is the observation that, in the 3D case, critical balance determines the relationship between perpendicular and parallel flux-tube dynamics.
The 3D system tends to have a larger kinetic-to-magnetic energy ratio than the 2D version (for the same initial Lundquist number $S_0$), mainly due to the flow arising in the initial stage of developing turbulence and the Alfv\'en waves propagating along the guide field. 
This renders the 3D system more turbulent than 2D. 
The $k^{-2}$ magnetic energy spectrum appears in 2D systems, caused by the sharp field reversals at thin current sheets. Although such current sheets are also present in the 3D simulations, this effect is masked by a $k^{-3/2}$ magnetic spectrum caused by the direct turbulent cascade.

This work lies exclusively in the (reduced) MHD framework. 
In many realistic astrophysical situations the magnetic field generation and early-stage evolution are most likely happening in collisionless plasmas without background magnetic fields. One may thus question the applicability of our results to such situations. While we do think that direct applicability may not be warranted, we also note that our analytical model for statistical flux tube mergers stems from the conservation of quantities that we would expect to remain conserved in the collisionless regime in the absence of background magnetic fields. We thus think that the ideas we have proposed here should form the basis of the corresponding collisionless model, which we defer to future study.

One may also wonder if the scaling laws of inverse energy transfer --- and magnetic reconnection as the process that enables such inverse transfer --- in systems with a strong guide field remain valid in isotropic systems (i.e., systems without a guide field), as investigated by~\citet{brandenburg2015nonhelical} and~\citet{zrake2014inverse}.
This is the subject of ongoing investigations~\citep{Bhat2019}; preliminary results suggest that is indeed the case.\\

\paragraph{\textbf{Acknowledgements.}}
This work was supported by NSF CAREER award No.~1654168 (MZ and NFL), NSF grants AST-1411879 and AST-1806084, and NASA ATP grants NNX16AB28G and NNX17AK57G (DAU).  
The authors thank Alexander Schekochihin, Annick Pouquet, Pallavi Bhat, Abhay Ram, Ryan White and Greg Werner for insightful discussions.
This research used resources of the MIT-PSFC partition of the Engaging cluster at the MGHPCC facility, funded by DOE award No.~DE-FG02-91-ER54109 and the National Energy Research Scientific Computing Center (NERSC)--a US Department of Energy Office of Science User Facility operated under Contract No. DE-AC02-05CH11231.
This work also used the Extreme Science and Engineering Discovery Environment (XSEDE), which is supported by National Science Foundation Grant No. ACI-1548562. This work used the XSEDE supercomputer Stampede2 at the Texas Advanced Computer Center (TACC) through allocation No. TG-PHY140041~\citep{towns2014xsede}.

\appendix
\section{Interaction between inverse magnetic transfer and ambient turbulence in the context of galactic magnetogenesis}
\label{sec:scale_estimation}
In this appendix we estimate the scale where the magnetic seed fields can be picked up and amplified by the turbulent dynamo. 

\begin{figure}
    \centering
    \includegraphics[width=0.6\textwidth]{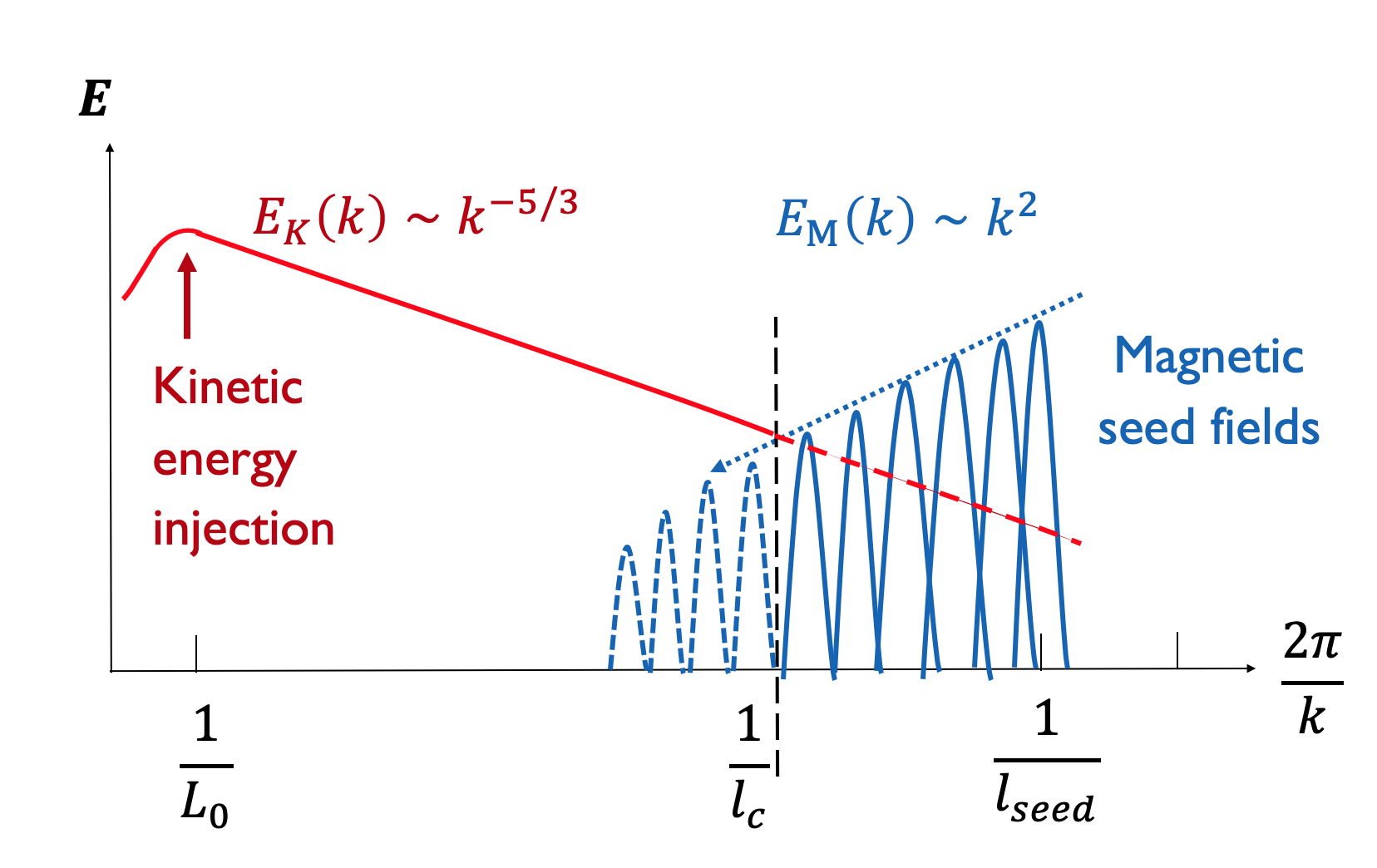}
    \caption{Schematic of kinetic energy spectrum and magnetic energy spectrum. Kinetic energy is injected at the large scale $L_0$ and cascade to smaller scales. Magnetic seed fields are generated at the small scale $l_{\rm seed}$ and inverse transfer to larger scales. }
    \label{fig:scheme_scale}
\end{figure}

We consider an unmagnetized hydro-turbulent astrophysical system where there is an asymptotically large scale separation between the injection of kinetic energy at large scales and the generation of magnetic seed fields at small scales.
A Kolmogorov cascade of kinetic energy is assumed, resulting in a $k^{-5/3}$ inertial range of kinetic energy spectrum $E_K(k)$, starting from the large driving scale $L_0$ down to the intermediate scales we are interested in. 
The evolution of magnetic seed fields is assumed to follow the scaling laws we derived in Section~\ref{sec:theory} [Eq.\eqref{eq:field_scalings} and Eq.\eqref{eq:length_scalings}].
We consider the seed fields to be generated at a small scale $l_{\rm seed}$, and described by a time-depending characteristic wavenumber $k(\tilde{t}) \sim \tilde{t}^{-1/2}$ as they merge and grow into larger scales.
The magnetic energy density of seed fields associated with its instantaneous characteristic wavenumber evolves as $E_M(\tilde{t}) \sim \tilde{t}^{-1}$, therefore, the magnetic energy density scales as $E_M(k) \sim k^2$. 
Fig.~\ref{fig:scheme_scale} shows a cartoon of the situation we have in mind: the kinetic and magnetic energy spectra resulting from the direct cascade and inverse transfer, respectively.
Kinetic energy dominates at the large scales while magnetic energy dominates at small scales. 
In this appendix, we provide an estimation of the critical scale $l_c$ where the kinetic and magnetic energy are comparable. At such scale, the ambient turbulent flow is dynamically important to the evolution of magnetic seed fields, and can act as a dynamo to amplify the magnetic field.

The kinetic energy at the critical scale $l_c$, $\mathcal{E}_K(l_c)$, and at the driving scale $L_0$, $\mathcal{E}_K(L_0)$,  are related by the $E_K(k)\sim k^{-5/3}$ spectrum, while the magnetic energy at $l_c$, $\mathcal{E}_M(l_c)$, and at the initial scale $l_{\rm seed}$, $\mathcal{E}_M(l_{\rm seed})$, are related by the $E_M(k) \sim k^2$ spectrum:
\begin{equation}
    \mathcal{E}_K(l_c) = \mathcal{E}_k(L_0)\left(\frac{l_c}{L_0}\right)^{5/3}, \qquad 
    \mathcal{E}_M(l_c) = \mathcal{E}_M(l_{\rm seed})\left(\frac{l_c}{l_{\rm seed}}\right)^{-2}. 
\end{equation}
The kinetic energy at the driving scale $L_0$ and magnetic energy at the initial scale $l_{\rm seed}$ can both be related to the plasma thermal pressure.
At the driving scale $L_0$, we have $\mathcal{E}_K(L_0) \sim M^2 P$, where $M \sim 1$ is the Mach number of the turbulence. At the seed field generation scale $l_{\rm seed}$, we have $\mathcal{E}_M(l_{\rm seed}) \sim \epsilon_B P$, where $\epsilon_B$ is a parameter of the saturation amplitude of the instability, and $\epsilon_B \sim 0.01$ if we consider the magnetic seed fields generated by the Weibel instability~\citep{silva2003interpenetrating}.

The critical scale $l_c$ calculated from the condition $\mathcal{E}_K(l_c) \sim \mathcal{E}_M(l_c)$ is:
\begin{equation}
\label{eq:equi_scale}
    l_c \sim l_{\rm seed}^{6/11}L_0^{5/11}\left(\frac{\epsilon_B}{M^2} \right)^{3/11},
\end{equation}
and the magnetic energy at this critical scale $l_c$ is:
\begin{equation}
    \mathcal{E}_M(l_c) \sim \left( \frac{l_{\rm seed}}{L_0} \right)^{10/11}P\epsilon_B^{5/11}M^{12/11}.
\end{equation}
From Eq.~\eqref{eq:equi_scale} we learn that the critical scale $l_c$ is close to the geometric mean between the driving large scale $L_0$ and the microscopic seed-field scale $l_{\rm seed}$.
We also note that the dependence of $l_c$ on the equipartition parameter $\epsilon_B$, which is determined by the mechanisms of seed field generation, is very weak.
The dependence of $l_c$ on the Mach number $M$, which is determined by various astrophysical events injecting kinetic energy into the system, is also weak.
Therefore, we think this estimation of the critical scale $l_c$ [Eq.~\eqref{eq:equi_scale}] should be a robust result, since it does not strongly depend on the astrophysical processes that generate the turbulence and magnetic seed fields, but on the generic features of the turbulent energy cascade, and the scaling laws for inverse energy transfer [Eq.\eqref{eq:field_scalings} and Eq.\eqref{eq:length_scalings}], which are derived based on the fundamental conservation laws.  

The scale of magnetic seed fields $l_{\rm seed}$ generated by the Weibel instability should be of the order of electron (or ion) skin depth $d_e$ (or $d_i$). 
For a typical galaxy, $L_0 \sim 30$ parsec $\sim 10^{20}$ cm, $d_e \sim 10^5$ cm and $d_i \sim 10^7$ cm~\citep{schekochihin2009astrophysical}. 
If we consider $l_{\rm seed} \sim d_e$, we obtain $l_c \sim 10^{11} \text{cm} \sim 10^6 d_e$ (if $l_{\rm seed} \sim d_i$, we obtain $l_c \sim 10^{12} \text{cm} \sim 0.1 \text{AU}$).
The strength of magnetic seed fields $B(l_{\rm seed}) \sim \mu G$ [about 3 $\mu G$ magnetic field corresponds to equipartition with the plasma pressure, where electron density $n_e \sim 1$ $\text{cm}^{-3}$ and temperature $T_e \sim 1$ eV are assumed in the interstellar medium (ISM)].
We thus obtain the strength of magnetic field at the critical scale $B(l_c) \sim B(l_{\rm seed}) (l_{\rm seed}/l_c) \sim 10^{-12}$ G for $l_{\rm seed} \sim d_e$ [$B(l_c) \sim 10^{-11}$ G for $l_{\rm seed} \sim d_i$].
We note that the strength of this magnetic field is significantly larger than that generated by the Biermann battery effect, while its scale is deep in the MHD range (since $l_c$ is approximately the geometric mean of the driving scale and plasma skin depth).

Our estimation is relevant for the problem of Galactic dynamo. 
Imagine we start with a completely unmagnetized and hydro-turbulent ISM, and a shock is launched through the ISM by an astrophysical event. 
The shock will produce microscopic filaments with typical radius of $10^7$ cm [on the order of ion skin depth~\citep{spitkovsky2008,Kato2008}], and with the amplitude of magnetic seed fields of the order of microgauss. 
These magnetic filaments would at first undergo a self evolution, without being affected by the turbulent flow at larger scales. 
During this stage, filaments will quickly merge with each other through magnetic reconnection, increasing their length scale and decreasing the strength of magnetic field.
This merging process would allow the filaments to deliver the magnetic field of the strength of $10^{-11}$ G to the scale of $10^{12}$ cm, where the turbulent flow is dynamically important.
This can be viewed as the starting point for the turbulent MHD dynamo. 
That is, the above gives an estimate for the scale and strength of the initial seed field for the Galactic dynamo problem.

\section{Current sheet stability  to ideal kink-type modes}
\label{sec:kink}
Our treatment and discussion of the dynamical evolution of an ensemble of flux tubes made no reference to the possible role of ideal kink-type instabilities.
In part this is justified by the fact that, in RMHD, a flux tube (more precisely, a screw pinch) with a circular cross-section is stable to the ideal internal kink mode~\citep{hazeltine2003plasma}.
One may wonder, however, if the highly elongated current sheets that arise as two flux tubes merge might themselves be kink-unstable. While it is unclear to us how to answer this question in full generality, it is possible to demonstrate stability for a simple one-dimensional current sheet described by a magnetic flux profile $\psi_0(x)$ and no flow, $\phi_0(x)=0$.

Consider, therefore, the ideal RMHD system of equations, and linearize them around the equilibrium $\psi_0(x)$, whose exact functional shape need not be specified for this proof.
We obtain:
\begin{align}
\label{eq:psi1}
&\gamma \psi_1-ik_y\phi_1\partial_x\psi_0=ik_zV_A\phi_1,\\
\label{eq:phi1}
&\gamma (\partial^2_x-k_y^2)\phi_1=[\psi_0,\nabla_\perp^2\psi_1]+[\psi_1,\nabla_\perp^2 \psi_0]+ik_zV_A(\partial_x^2-k_y^2)\psi_1.
\end{align}

The combination of these two relations yields the following equation for the perturbed stream function $\phi_1$:
\begin{equation}
\label{eq:master}
   \left[ \gamma^2+(k_zV_A+k_y\partial_x\psi_0)^2 \right](\partial_x^2-k_y^2)\phi_1+2k_y(k_zV_A+k_y\partial_x\psi_0)\partial_x^2\psi_0 \partial_x \phi_1 = 0.
\end{equation}
It can be shown that, as expected, Eq.~\eqref{eq:master} is self-adjoint, since it is derived from the ideal MHD equations possessing the self-adjoint properties.
That is, it can be written in the form
\begin{align}
     \mathcal{L} \phi_1 \equiv \partial_x \left[f(x)\partial_x \phi_1 \right]+g(x) \phi_1=0,
\end{align}
where $\mathcal{L}$ is a self-adjoint operator and 
\begin{align}
    \label{eq:fx}
   & f(x)=\gamma^2+(k_zV_A+k_y\partial_x \psi_0)^2,\\
   & g(x)=-k_y^2f(x).
\end{align}

Due to the properties of self-adjoint operators, the eigenvalues $\gamma$ can only be either real or imaginary. If $\gamma^2 < 0$, the modes are purely oscillatory, and if $\gamma^2 > 0$, the modes are purely growing or damped. 
The self-adjoint operator $\mathcal{L}$ satisfies $\int_{-\infty}^\infty \phi_1^* \mathcal{L} \phi_1 dx=0$. 
Therefore Eq.~\eqref{eq:master} yields:
\begin{equation}
\begin{aligned}
\int_{-\infty}^\infty \phi_1^*\left[\partial_x\left(f(x)\partial_x \phi_1\right)-k_y^2f(x)\phi_1\right]dx=0.
\end{aligned}
\end{equation}
Using the boundary condition $\phi_1 \rightarrow 0$ as $x \rightarrow \pm \infty$, we obtain:
\begin{equation}
\begin{aligned}
\label{eq:int1}
\int_{-\infty}^\infty \left(f(x) \partial_x \phi_1^* \partial_x \phi_1+k_y^2f(x)\phi_1^*\phi_1\right)dx = 0.
\end{aligned}
\end{equation}
Equivalently, Eq.~\eqref{eq:int1} can be written as an expression for $\gamma^2$:
\begin{equation}
\begin{aligned}
\gamma^2 = -\frac{\int_{-\infty}^\infty (k_zV_A+k_y\partial_x \psi_0)^2 \left( \partial_x \phi_1^* \partial_x \phi_1+k_y^2\phi_1^*\phi_1\right)dx }{\int_{-\infty}^\infty \left( \partial_x \phi_1^* \partial_x \phi_1+k_y^2\phi_1^*\phi_1\right)dx }.
\end{aligned}
\end{equation}
This expression makes the fact that $\gamma^2<0$ explicit for any nontrivial solution $\phi(x)$. 
I.e., any admissible modes are purely oscillatory: essentially, shear Alfv\'en waves propagating in a nontrivial magnetic field. The specific values of $\gamma^2$ will depend on the functional form of $\psi_0(x)$, but not their sign. 

Based on this analysis, we conclude that two representative structures of the system that we study in this paper --- cylindrical flux tubes and the current sheets between them --- are stable to ideal kink-type modes.
This is not, of course, a proof that the kink instability plays no role in the evolution of our system. 
However, in addition, we call attention to the fact that, at least at the twiddle-algebra level, the critical balance condition, which our system does seem to obey rather well, is the condition for kink stability. 
Altogether, we think these are reasonably convincing arguments for ignoring ideal kink-type modes in our analysis.

\section{Inverse energy transfer of random magnetic fields}
\label{sec:noise}
The inverse transfer of magnetic energy and the formation of large-scale structures in a system consisting of volume-filling flux tubes can be explained by our hierarchical model based on the merger of flux tubes, as we demonstrate in the main part of this paper.
However, it is possible that this fundamental physical process --- merger of magnetic structures through magnetic reconnection --- can be a more general phenomenon. 
It is well known that in 3D MHD turbulence an inverse helicity cascade and the corresponding inverse energy transfer occurs due to the conservation of (non-zero) net magnetic helicity~\citep{pouquet_1978,christensson2001}. Recently, however, \citet{brandenburg2015nonhelical} and \citet{zrake2014inverse} numerically demonstrated that an inverse energy transfer exists even when the net magnetic helicity is zero, in 3D decaying non-relativistic and relativistic turbulence respectively.
A careful parameter study of this problem can be found in 
\citet{Reppin2017}.
\citet{Bhat2019} argue that this non-helical inverse transfer phenomenon might be explained by the merger of magnetic structures in a qualitatively similar way to the process that we have described in this paper and in~\citetalias{zhou2019magnetic}.

To further explore and substantiate this idea – and, more generally, to demonstrate that the inverse transfer that we observe is a generic phenomenon, and not something that depends on the specific form of the initial magnetic field --- we present in this appendix a simulation which differs from the ones analyzed in the main body of the paper in that the initial condition is instead a random magnetic field. 
Specifically, we initialize a Gaussian-random magnetic field, with the perpendicular magnetic spectrum narrowly peaked around $k_{\perp,0}=20$.
In the parallel direction we impose sinusoidal perturbations using the modes with the 16 longest wavelengths and random phases.
The simulation is performed with $512^2 \times 256$ grid points, and we have used hyper-dissipation rather than regular (Laplacian) resistivity and viscosity.

\begin{figure}
\includegraphics[width=\textwidth,clip]{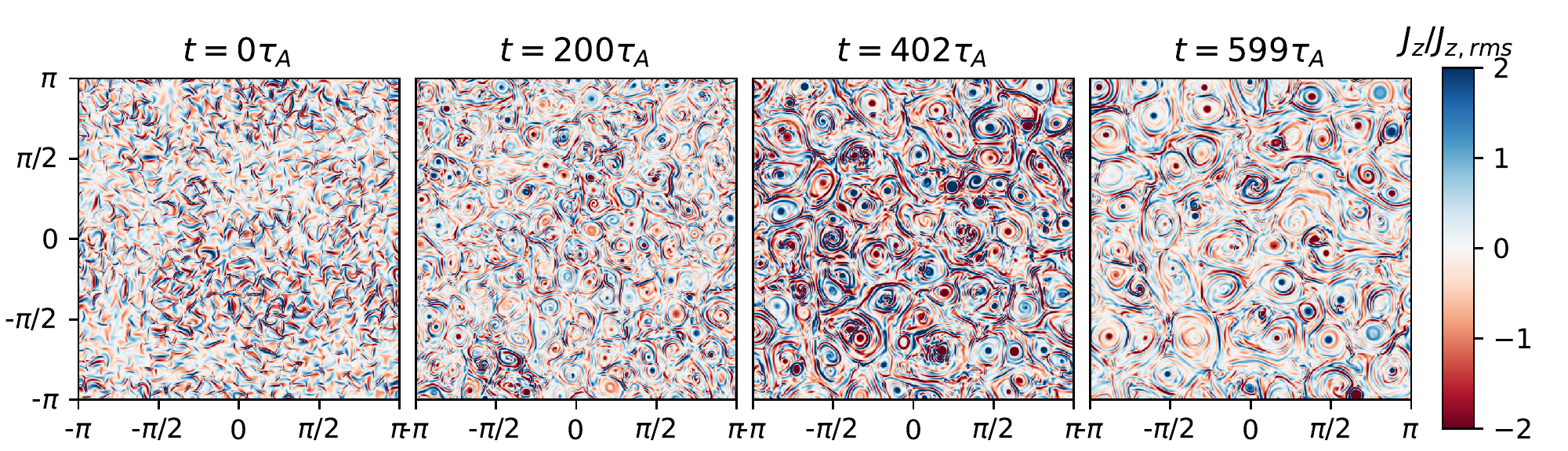}
\caption{Current density $J_z$ (normalized by its root mean square value) at various times on the $xy$ plane at $z=0$ for a run starting with a random magnetic field.}
\label{fig:2d_vis_noise}
\end{figure}

\begin{figure}
    \centering
    \includegraphics[width=0.45\textwidth]{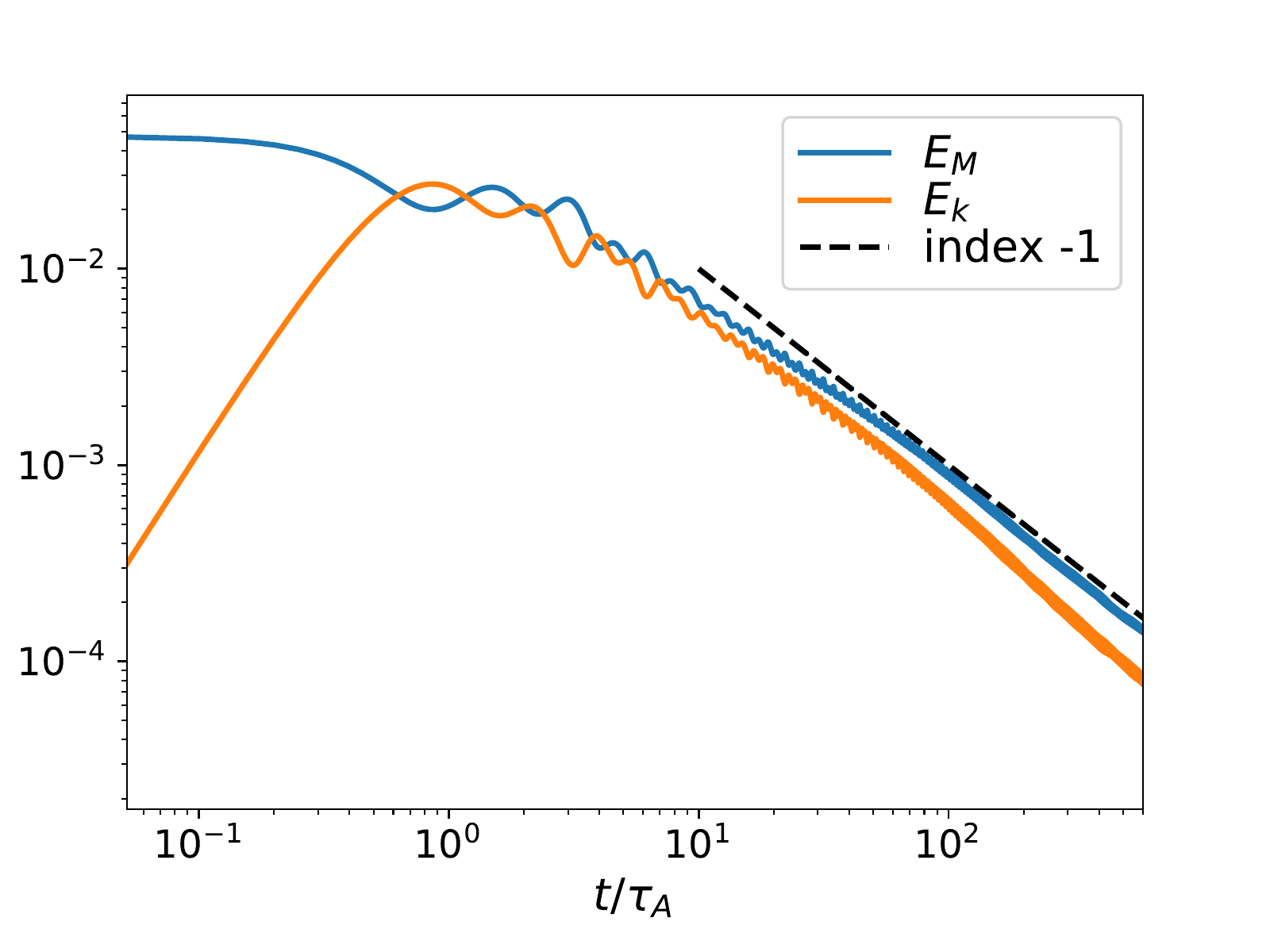}
    \caption{Time evolution of total magnetic ($E_M$) and kinetic ($E_K$) energies for the run starting with random magnetic field.}
    \label{fig:energy_decay_noise}
\end{figure}

\begin{figure}
    \centering
    \includegraphics[width=0.45\textwidth]{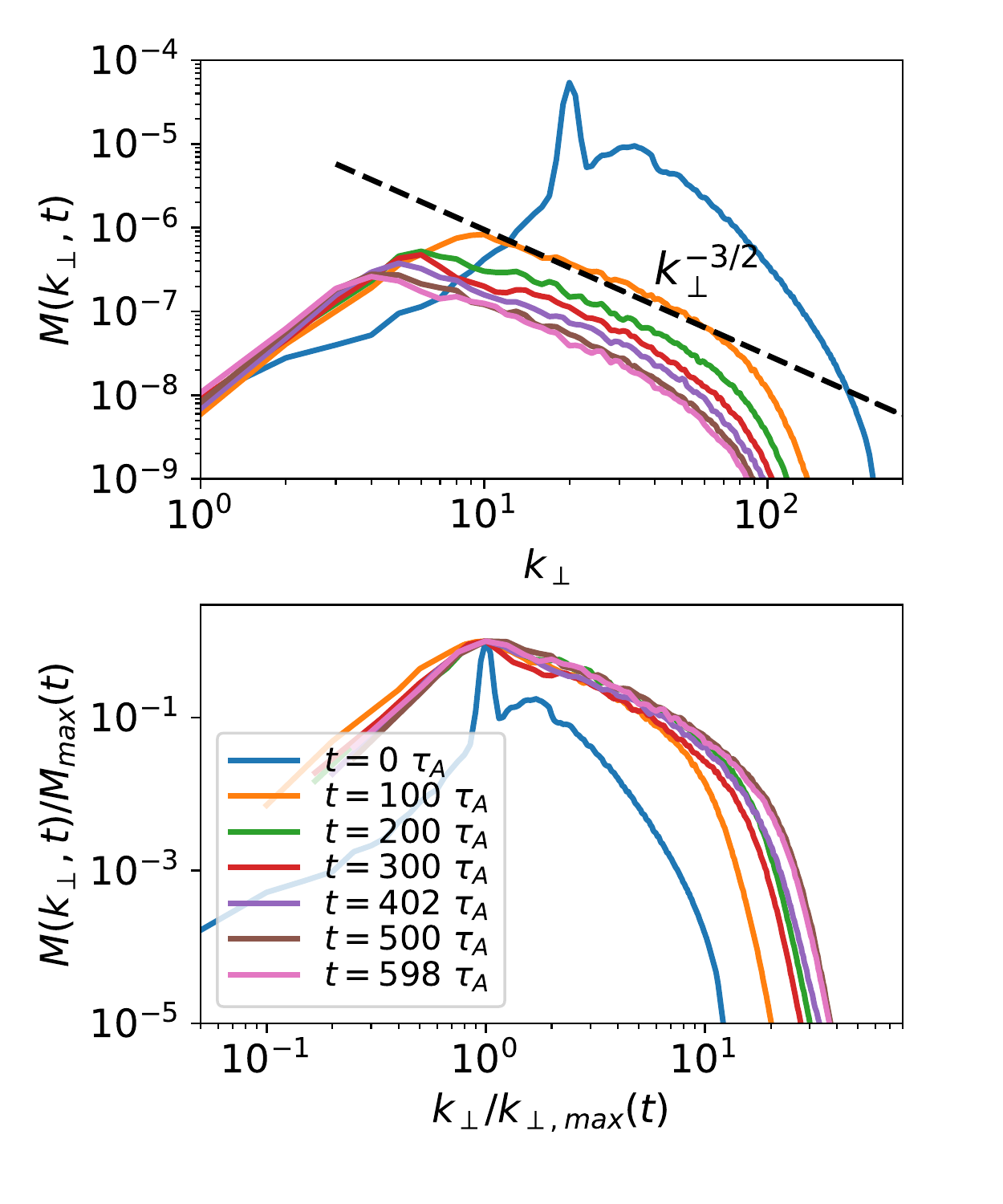}
    \caption{Raw (top) and normalized (bottom) perpendicular magnetic power spectra, for a simulation starting with random magnetic fields. A $k_\perp^{-3/2}$ slope is shown for reference.}
    \label{fig:mag_spec_noise}
\end{figure}

From the visualizations of current density shown in Fig.~\ref{fig:2d_vis_noise}, we see the emergence of larger structures from the initial small-scale random magnetic field.
The behaviors of the magnetic energy decay and of the perpendicular magnetic spectrum are similar to the runs with the flux-tube setup, and are consistent with the predictions of our hierarchical model. 
The time evolution of magnetic energy (Fig.~\ref{fig:energy_decay_noise}) shows the $t^{-1}$ scaling.
In Fig.~\ref{fig:mag_spec_noise} we show the time evolution of the perpendicular magnetic power spectrum $M(k_\perp,t)$. As the system evolves, a $k_\perp^{-3/2}$ inertial range is developed, and the spectra are observed to migrate to larger scales (Fig.~\ref{fig:mag_spec_noise}, top panel), demonstrating the inverse energy transfer.
When normalized as described in Section~\ref{sec:spectra}, the spectra roughly overlap (Fig.~\ref{fig:mag_spec_noise} bottom panel), indicating the self-similar evolution of the system, as predicted by our hierarchical model in Section~\ref{sec:selfsimilar-scaling}.

The similarities between this simulation and those reported in the main section of this paper strongly support the notion that  non-helical inverse transfer enabled by magnetic reconnection is a generic phenomenon.

\makeatletter
\let\jnl@style=\rm
\def\ref@jnl#1{{\jnl@style#1}}
\def\aj{\ref@jnl{AJ}}                   
\def\actaa{\ref@jnl{Acta Astron.}}      
\def\araa{\ref@jnl{ARA\&A}}             
\def\apj{\ref@jnl{ApJ}}                 
\def\apjl{\ref@jnl{ApJ}}                
\def\apjs{\ref@jnl{ApJS}}               
\def\ao{\ref@jnl{Appl.~Opt.}}           
\def\apss{\ref@jnl{Ap\&SS}}             
\def\aap{\ref@jnl{A\&A}}                
\def\aapr{\ref@jnl{A\&A~Rev.}}          
\def\aaps{\ref@jnl{A\&AS}}              
\def\azh{\ref@jnl{AZh}}                 
\def\baas{\ref@jnl{BAAS}}               
\def\bac{\ref@jnl{Bull. astr. Inst. Czechosl.}} 
\def\caa{\ref@jnl{Chinese Astron. Astrophys.}}
\def\cjaa{\ref@jnl{Chinese J. Astron. Astrophys.}}
\def\icarus{\ref@jnl{Icarus}}           
\def\jcap{\ref@jnl{J. Cosmology Astropart. Phys.}}
\def\jrasc{\ref@jnl{JRASC}}             
\def\memras{\ref@jnl{MmRAS}}            
\def\mnras{\ref@jnl{MNRAS}}             
\def\na{\ref@jnl{New A}}                
\def\nar{\ref@jnl{New A Rev.}}          
\def\pra{\ref@jnl{Phys.~Rev.~A}}        
\def\prb{\ref@jnl{Phys.~Rev.~B}}        
\def\prc{\ref@jnl{Phys.~Rev.~C}}        
\def\prd{\ref@jnl{Phys.~Rev.~D}}        
\def\pre{\ref@jnl{Phys.~Rev.~E}}        
\def\prl{\ref@jnl{Phys.~Rev.~Lett.}}    
\def\pasa{\ref@jnl{PASA}}               
\def\pasp{\ref@jnl{PASP}}               
\def\pasj{\ref@jnl{PASJ}}               
\def\rmxaa{\ref@jnl{Rev. Mexicana Astron. Astrofis.}}
\def\qjras{\ref@jnl{QJRAS}}             
\def\skytel{\ref@jnl{S\&T}}             
\def\solphys{\ref@jnl{Sol.~Phys.}}      
\def\sovast{\ref@jnl{Soviet~Ast.}}      
\def\ssr{\ref@jnl{Space~Sci.~Rev.}}     
\def\zap{\ref@jnl{ZAp}}                 
\def\nat{\ref@jnl{Nature}}              
\def\iaucirc{\ref@jnl{IAU~Circ.}}       
\def\aplett{\ref@jnl{Astrophys.~Lett.}} 
\def\apspr{\ref@jnl{Astrophys.~Space~Phys.~Res.}}     
\def\bain{\ref@jnl{Bull.~Astron.~Inst.~Netherlands}}   
\def\fcp{\ref@jnl{Fund.~Cosmic~Phys.}}  
\def\gca{\ref@jnl{Geochim.~Cosmochim.~Acta}}   
\def\grl{\ref@jnl{Geophys.~Res.~Lett.}} 
\def\jcp{\ref@jnl{J.~Chem.~Phys.}}      
\def\jgr{\ref@jnl{J.~Geophys.~Res.}}    
\def\jqsrt{\ref@jnl{J.~Quant.~Spec.~Radiat.~Transf.}}   
\def\memsai{\ref@jnl{Mem.~Soc.~Astron.~Italiana}}    
\def\nphysa{\ref@jnl{Nucl.~Phys.~A}}   
\def\physrep{\ref@jnl{Phys.~Rep.}}   
\def\physscr{\ref@jnl{Phys.~Scr}}   
\def\planss{\ref@jnl{Planet.~Space~Sci.}}   
\def\procspie{\ref@jnl{Proc.~SPIE}}   

\let\astap=\aap
\let\apjlett=\apjl
\let\apjsupp=\apjs
\let\applopt=\ao
\makeatother
\bibliographystyle{jpp}

\bibliography{ref}

\end{document}